\pdfoutput=1
\documentclass[aps,cha,preprint,groupedaddress,nofootinbib]{revtex4-1}
\usepackage[]{natbib}
\usepackage{amsmath, amssymb, amsthm, url}
\usepackage{graphicx}
\usepackage[table]{xcolor}  
\usepackage[normalem]{ulem} 
\usepackage{dcolumn}
\usepackage{setspace}
\usepackage{booktabs}
\newcolumntype{d}[1]{D{.}{.}{#1}}

\usepackage{booktabs}

\newcolumntype{x}[1]{>{\centering\arraybackslash\hspace{0pt}}m{#1}}
\newcolumntype{.}{D{.}{.}{3.0}}

\begin{document}


\title{Mathematical model of gender bias and homophily in professional hierarchies}
\spacing{1}


\author{Sara M.~Clifton}
\email{smc567@illinois.edu}
\affiliation{Department of Mathematics, University of Illinois at Urbana-Champaign, Urbana, Illinois 61801, USA}

\author{Kaitlin Hill}
\affiliation{School of Mathematics, University of Minnesota, Minneapolis, Minnesota 55455, USA}

\author{Avinash J.~Karamchandani}
\affiliation{Department of Engineering Sciences and Applied Mathematics, Northwestern University, Evanston, Illinois 60208, USA}

\author{Eric A.~Autry}
\affiliation{Department of Mathematics, Duke University, Durham, North Carolina 27708, USA}

\author{Patrick McMahon}
\affiliation{Department of Electrical and Computer Engineering, University of Illinois at Urbana-Champaign, Urbana, Illinois 61801, USA}

\author{Grace Sun}
\affiliation{Department of Physics, University of Illinois at Urbana-Champaign, Urbana, Illinois 61801, USA  \vspace{1.5cm}}



\begin{abstract} 
Women have become better represented in business, academia, and government over time, yet a dearth of women at the highest levels of leadership remains. Sociologists have attributed the leaky progression of women through professional hierarchies to various cultural and psychological factors, such as self-segregation and bias. Here, we present a minimal mathematical model that reveals the relative role that bias and homophily (self-seeking) may play in the ascension of women through professional hierarchies. Unlike previous models, our novel model predicts that gender parity is not inevitable, and deliberate intervention may be required to achieve gender balance in several fields. To validate the model, we analyze a new database of gender fractionation over time for 16 professional hierarchies. We quantify the degree of homophily and bias in each professional hierarchy, and we propose specific interventions to achieve gender parity more quickly.
\end{abstract}

\maketitle




\section{Introduction}
A professional hierarchy is a field in which an employee enters at a designated low level and gradually moves up the ranks. For instance, large businesses have interns through CEOs, hospitals have residents through head physicians, and academic institutions have undergraduates through full professors. Over time, women have generally become better represented in many industries (e.g., \cite{luckenbill2002educational, kaye2007progress, terjesen2009women,farrell2005additions}), but women are still poorly represented at the highest levels of most professional hierarchies (e.g., \cite{nelson2003national, carr2015inadequate,shapiro2015middle}). This has been called the `leaky pipeline' effect.

Countless factors have been proposed to explain this so-called `leaky pipeline' effect: family responsibilities \cite{eagly2012women}, different professional interests between the genders \cite{konrad2000sex,ceci2010sex,van2004academic}, biological differences \cite{sapienza2009gender}, unconscious bias in the workplace \cite{easterly2011conscious,lee2005unconscious}, laws restricting gender discrimination \cite{chacko1982women,sape1971title}, societal gender roles \cite{britton2000epistemology,blackburn1995measurement,dean2015work}, and other entrenched cultural or psychological factors. Many of these qualitative theories require an implicit assumption that men and women make fundamentally different decisions, either as a result of biological differences or social indoctrination. 

Some quantitative models have attempted to study the ascension of women through certain fields without relying on intrinsic differences between the sexes. Shaw and Stanton \cite{shaw2012leaks} calculate the `inertia' of women through several academic hierarchies, and find that gender differences play a diminishing role in promotion over time. Holman et al. \cite{holman2018gender} present the first quantitative model to our knowledge that attempts to predict the time required to reach gender parity in academic STEM fields, with estimates as high as several centuries in some disciplines. Their model assumes logistic growth to gender parity of the proportion of women in senior and junior academic roles (as estimated by last and first authorship on research papers, respectively). Although logistic growth and eventual gender parity is a reasonable assumption for their phenomenological model, we create a mechanistic model to examine the relative impact of two major sociological factors, homophilic (self-seeking) instincts and gender bias, on the progression of women through professional hierarchies. We find that gender parity is not guaranteed, and gender fractionation may never settle to an equilibrium.

\section{Model}
Broadly speaking, two classes of people influence the ascension of individuals through a professional hierarchy: people at lower levels choose to apply for higher positions, and people at higher levels choose to promote applicants into the next level. People at higher levels affect the promotion of individuals through their hiring biases, while the decisions to apply for promotion made by those at lower levels are affected by their own homophilic tendencies. 

Women in hierarchical professions tend to be promoted more slowly than men, even when accounting for differing productivity and attrition, indicating that gender is a salient factor in the hiring process \cite{tesch1995promotion,heilman2001description,eagly2007women,kumra2008study,winkler2000faculty}. If gender is the determining factor when deciding between equally qualified candidates, we will say that a gender bias exists. We define gender bias as all conscious or unconscious decisions made by the employer during the hiring process that are affected by the gender of the applicant. For simplicity, we will assume that gender-based hiring bias is constant across all hierarchy levels (i.e., employers will uniformly reduce or enhance female candidates' relative chance of promotion at all levels).

Gendered differences in promotion also depend on gender differences in the applicant pool due to individuals self-selecting, consciously or unconsciously, whether or not to submit an application. When gender is a salient factor in deciding whether or not to submit an application, we will assume that such decisions are based largely on a homophilic instinct. In other words, when an individual considers whether or not to apply for a promotion, he or she looks at the demographics of those working at the above level and evaluate whether or not they `belong' in that higher level. While this assumption may seem simplistic, many studies show that people unconsciously self-segregate based on gender from a early age \cite{maccoby1987gender,alexander1994gender,moller1996antecedents,powlishta1995gender}. In fact, perceptions of gendered jobs perpetuate much of the occupational gender segregation we see today \cite{miller2004occupational,pan2015gender,ludsteck2014impact,alonso2012extent,bender2005job}.

With the goal of understanding the relative roles that bias and homophily (self-segregation) play in the ascension of women through professional hierarchies, we derive a minimal mathematical model that incorporates both forces. To introduce the model, we begin with a simple example.

\subsection{Example}
\begin{figure}[htb]
  \begin{center}
    \includegraphics[width=0.65\textwidth]{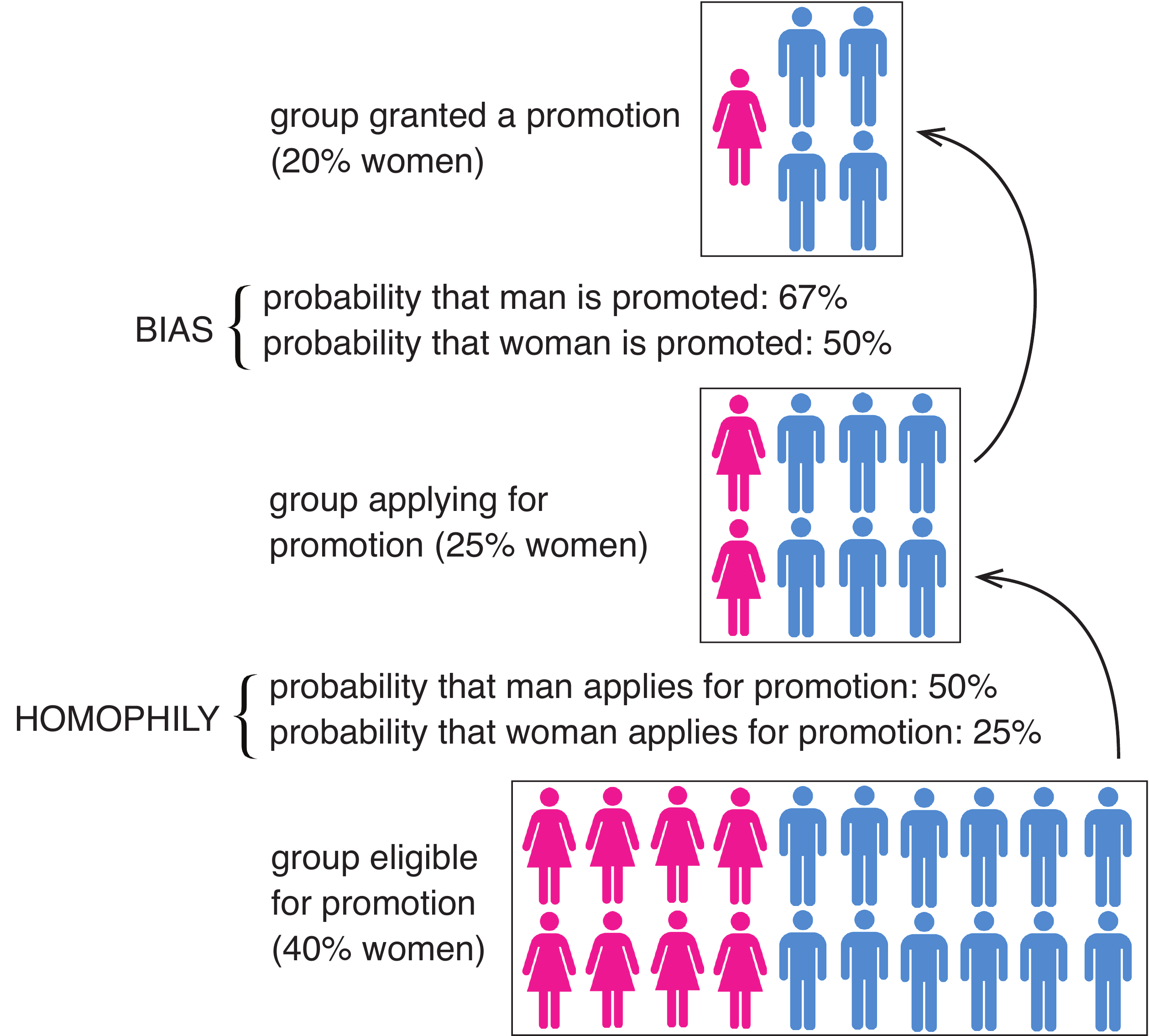}
  \end{center}
  \caption{Example of a potential decision process between two levels in a professional hierarchy.}
  \label{fig:example}
\end{figure}

Consider the decision process that occurs during the transition between two levels in a professional hierarchy (Figure \ref{fig:example}). Suppose the lower level is 40\% women, and gender is not a factor in eligibility for promotion; then the group eligible for promotion is also 40\% women. If women are not well-represented in the higher level, then women may not feel as comfortable applying for promotion as men. To be clear, we do not suppose that women are intrinsically less likely to apply for promotion; rather we assume that the gender demographics in the upper level affect both men and women's feeling of belonging (homophily) in the upper level.

\begin{figure}[htb]
  \begin{center}
    \includegraphics[width=0.7\textwidth]{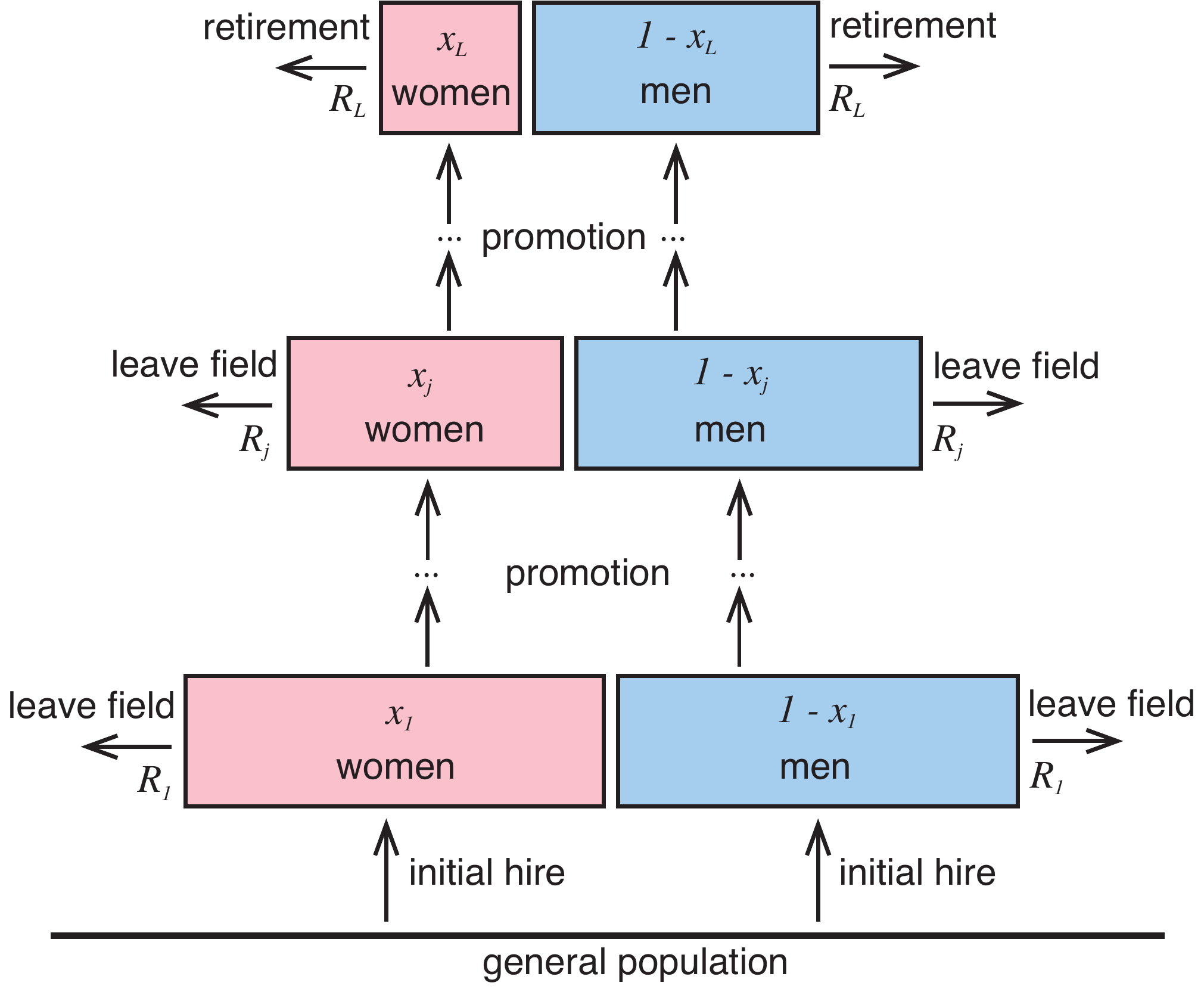}
  \end{center}
  \caption{Schematic of an $L$-level hierarchy. The $j$th level in the hierarchy has a certain fraction women $x_j$, and people retire or leave the field from each level at a rate $R_j$. The general population is assumed to be $1/2$ women at all times.}
  \label{fig:schematic}
\end{figure}

Say men are twice as likely to apply for promotion due to these homophilic instincts. Then the applicant pool will shrink to 25\% women. If no bias towards or against women exists in hiring, then 25\% of those granted promotion will be women. However, if women are slightly less likely than men to be granted promotion due to bias, then the fraction of women hired will shrink again. We assume this decision process occurs between all levels in a professional hierarchy. The schematic in Figure \ref{fig:schematic} is a visualization of a generic hierarchy.

\subsection{Model derivation}
\begin{figure}[htb]
  \begin{center}
    \includegraphics[width=0.75\textwidth]{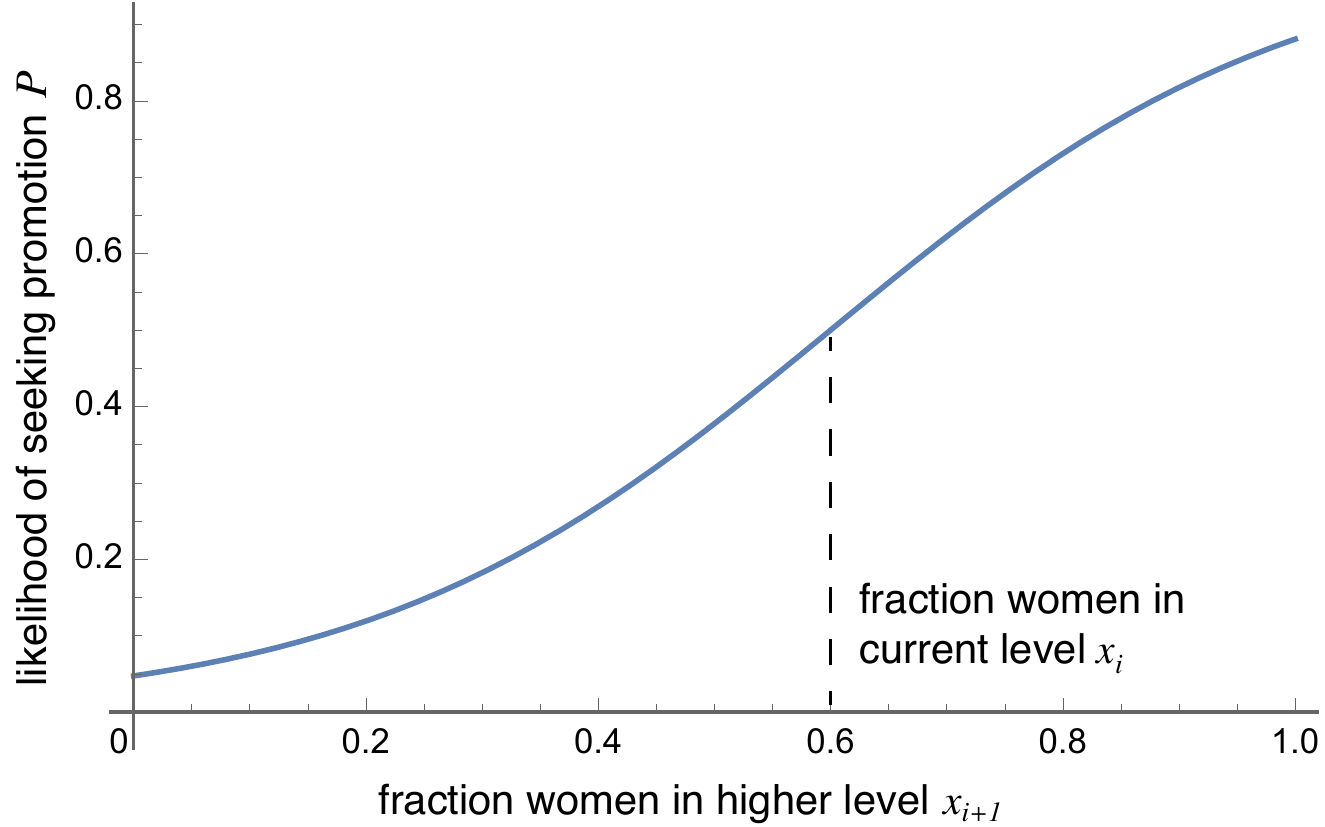}
  \end{center}
  \caption{An example of the probability that a woman seeks promotion, dependent on the demographics of the level to which she is applying. In this example, a woman is more likely to apply for promotion if there are more women in the level above her. The probability changes most rapidly around the demographic split she is most accustomed to, the gender split in her current position.}
  \label{fig:P}
\end{figure}

We begin by assuming that the probability $P(u,v)$ of seeking promotion to the next level is a function of the fraction of people at the upper level who share the applicant's gender, $u$, and the fraction of like-gendered individuals in the applicant's current level, $v$. There exists a `one-third hypothesis' that supports the anecdotal evidence that an individual feels comfortable in a group environment when at least 30\% of the members share the individual's demographic status \cite{engelmann1967communication,srikantan1968curious}. To our knowledge, this hypothesis has not been rigorously tested in the real world, so we allow $P$ to take a more flexible form. Specifically, we suppose that the threshold of comfort may depend on the environment in which a person currently resides. We also assume that the threshold does not delineate an instantaneous switch from 0\% comfort to 100\% comfort; instead the comfort level may gradually change around that threshold. One simple function that captures this behavior is the sigmoid
\begin{equation} \label{eq:P}
	P(u,v) = \frac{1}{1+ e^{-\lambda(u-v)}},
\end{equation}
\noindent where $u$ is the fraction of like-gendered individuals in the level above, $v$ is the fraction of like-gendered individuals in the current level, and $\lambda$ is the strength of the homophilic tendency. This function need not be a literal probability because only the relative likelihood of applying for promotion is relevant. Because we choose to not include inherent gender difference in the model, we assume that this function applies to both men and women. See Figure \ref{fig:P} for a sketch of this homophily function.

Given this probability $P(u,v)$ of seeking promotion, the fraction of women in the applicant pool is 
\begin{align} \label{eq:f0}
f_0(u,v) = \frac{v P(u,v)}{v P(u,v) + (1-v) P(1-u,1-v)},
\end{align}
\noindent where $u$ is the fraction of women in the higher level and $v$ is the fraction of women in the current level.

In addition to self-segregation dynamics, hiring bias towards or against women will change the proportion of female applicants who are promoted. We incorporate this constant bias $b$ as the female fraction of those promoted if the applicant pool has an equal number of men and women. For instance, a bias $b$ exceeding $\tfrac{1}{2}$ would imply that women are favored disproportionately, and a bias less than $\tfrac{1}{2}$ suggests that men are favored. The fraction of women promoted to the next level is then
\begin{align}
f(u,v; b) = \frac{b \, v \, P(u,v)}{b \, v \, P(u,v) + (1-b) (1-v) P(1-u,1-v)}.
\end{align}
\noindent This is not the only way to incorporate bias, but it is a simple way to ensure that bias does not leave vacancies or induce the promotion of those who have not applied. As an example, a naive choice to incorporate bias would be $f(u,v; b)=b f_0(u,v)$, where $b<1$ indicates bias against women. However, this choice permits $f>1$ if $b$ or $f_0$ are sufficiently large. 

Because professional hierarchies are frequently competitive, with each level smaller than the level below it, we assume that all vacancies will be filled. The vacancies are created by individuals who are promoted to the next level, those leaving the field at a particular level, or those retiring from the top level. The change in the number of women at each level, $x_j N_j$, is 
\begin{equation}  \begin{aligned}
	\frac{\mathrm{d}}{\mathrm{d} t} (x_L N_L) =& \, R_L N_L\, f\left(x_L,x_{L-1}; b \right) - R_L N_L \, x_L \\
	\frac{\mathrm{d}}{\mathrm{d} t} (x_j N_j) =& \left( \sum_{k=j}^{L} R_k N_k \right) f\left(x_j,x_{j-1}; b\right) -R_j N_j x_j  \\
&- \left( \sum_{k=j+1}^{L} R_k N_k \right) f\left(x_{j+1}, x_{j}; b\right) \,\,\,\,\,\,\, \text{for } 1< j < L \\
	\frac{\mathrm{d}}{\mathrm{d} t} (x_1 N_1) =& \left( \sum_{k=1}^{L} R_k N_k \right) f(x_1, \tfrac{1}{2}; b) -R_1 N_1 x_1 - \left( \sum_{k=2}^{L} R_k N_k \right) f\left(x_{2}, x_{1}; b\right),
\label{eq:derive}
\end{aligned} \end{equation}
\noindent where $L$ is the number of levels in the hierarchy, $x_j$ is the fraction of people in level $j$ who are women, $N_{j}$ is the number of people in the $j$th level, $R_j$ is the retirement/leave rate at the $j$th level, $\sum_{k=j+1}^{L} R_k N_k$ is the total number of retiring people above the $j$th level, $b$ is the bias parameter, and $f(\,\cdot \,)$ is the fraction of people promoted to the next level who are women. Because it may not be intuitive that the change in the number of women at lower levels depends on the total number of retiring people above the level, we provide a simple example to illustrate this feature in the Appendix.

We normalize system \eqref{eq:derive} by dividing each equation by the number of people retiring/leaving the level (i.e., $R_j N_j$):
\begin{equation}  \begin{aligned}
\frac{1}{R_L}\frac{\mathrm{d} x_L}{\mathrm{d} t} &= \overbrace{f\left(x_L,x_{L-1}; b\right)}^{\substack{\text{promoted from} \\ \text{lower level}}} - \overbrace{x_L}^{\substack{\text{retire out} \\ \text{of level}}} \\
\frac{1}{R_j}\frac{\mathrm{d} x_j}{\mathrm{d} t} &= (1+r_j) f\left(x_j,x_{j-1}; b \right) - x_j - r_j f\left(x_{j+1}, x_{j}; b\right) \,\,\,\,\,\,\, \text{for } 1< j < L \\
\frac{1}{R_1}\frac{\mathrm{d} x_1}{\mathrm{d} t} &= \underbrace{(1+r_1) f(x_1,\tfrac{1}{2}; b)}_{\substack{\text{hired from} \\ \text{general pool}}} - \underbrace{x_1}_{\substack{\text{leave} \\ \text{field}}} - \underbrace{r_1 f\left(x_{2}, x_{1}; b\right)}_{\substack{\text{promoted to} \\ \text{next level}}},
\label{eq:full}
\end{aligned} \end{equation}
\noindent where $r_j$ is the ratio of the total retiring people above the $j$th level to the retiring people in the $j$th level. Algebraically, this ratio is $ r_j = \sum_{k=j+1}^{L} R_k N_k/R_j N_j$. Note that this system can be condensed into one line by taking $r_L = 0$ and $x_0 = 1/2$. Refer to Table \ref{tab:parameters} for descriptions of the model variables and parameters.

\begin{table}[htb]
\begin{center}
  \begin{tabular}{| p{1.5cm}  p{10cm} |} \hline
    Variable & Meaning \\ \hline 
    $x_j$ & fraction of people in the $j$th level who are women  \\
    $L$ & number of levels in hierarchy \\
    $R_j$ & retirement/leave rate at the $j$th level \\ 
    $N_j$ & number of people in the $j$th level \\
    $r_j$ & ratio of the total retiring people above the $j$th level to the retiring people in the $j$th level $\left(\sum_{k=j+1}^{L} R_k N_k/R_j N_j\right)$ \\ 
    $P(\,\cdot\,)$ & likelihood of seeking promotion \\ 
    $f(\,\cdot\,)$ & fraction of people promoted to next level who are women \\
    $b$ & bias towards or against women ($b=1/2$ is no bias) \\ 
    $\lambda$ & strength of homophilic tendency \\ 
    \hline
\end{tabular}
     \caption{Model variables and parameters for \eqref{eq:full}.}
  \label{tab:parameters}
\end{center} 
\end{table}

\subsection{Null model}
Consider a null model with no hiring bias or homophily. In model \eqref{eq:full}, this would imply bias $b=\tfrac{1}{2}$ and the likelihood of seeking promotion $P$ is constant ($\lambda=0$). The model then reduces to the linear system 
\begin{equation}
\frac{1}{R_j} \frac{\mathrm{d} x_j}{\mathrm{d} t} = (1+r_j) (x_{j-1} - x_j )  \,\,\,\,\,\,\, \text{for } 1\le j \le L.
\label{eq:null}
\end{equation}
The only steady state is $\{ x_j^{*} \} = \{ \tfrac{1}{2} \}$. The Jacobian of the system evaluated at this state yields all real, negative eigenvalues. Therefore, $\{ x_j \} = \{ \tfrac{1}{2} \}$ is a stable sink of the null model. In other words, without bias or homophily, each level in the hierarchy will directly converge to equal gender representation, as seen in the model by Holman et al. \cite{holman2018gender}. The rate of convergence to parity for each level depends on the eigenvalues of the system: $\lambda_j=-R_j(1+r_j),$ for $j=1,\ldots,L$. The eigenvalues depend only on the level sizes $N_j$ and leave rates $R_j$. The convergence time to parity for the whole system is then given by $1/\min_j\{ R_j(1+r_j) \},$ the characteristic timescale of the system. 

See 
the Appendix for a more complete discussion of analysis of the null model. Figure \ref{fig:null} shows convergence to gender parity in a hypothetical academic hierarchy.

\begin{figure}[htb]
  \begin{center}
    \includegraphics[width=0.6\textwidth]{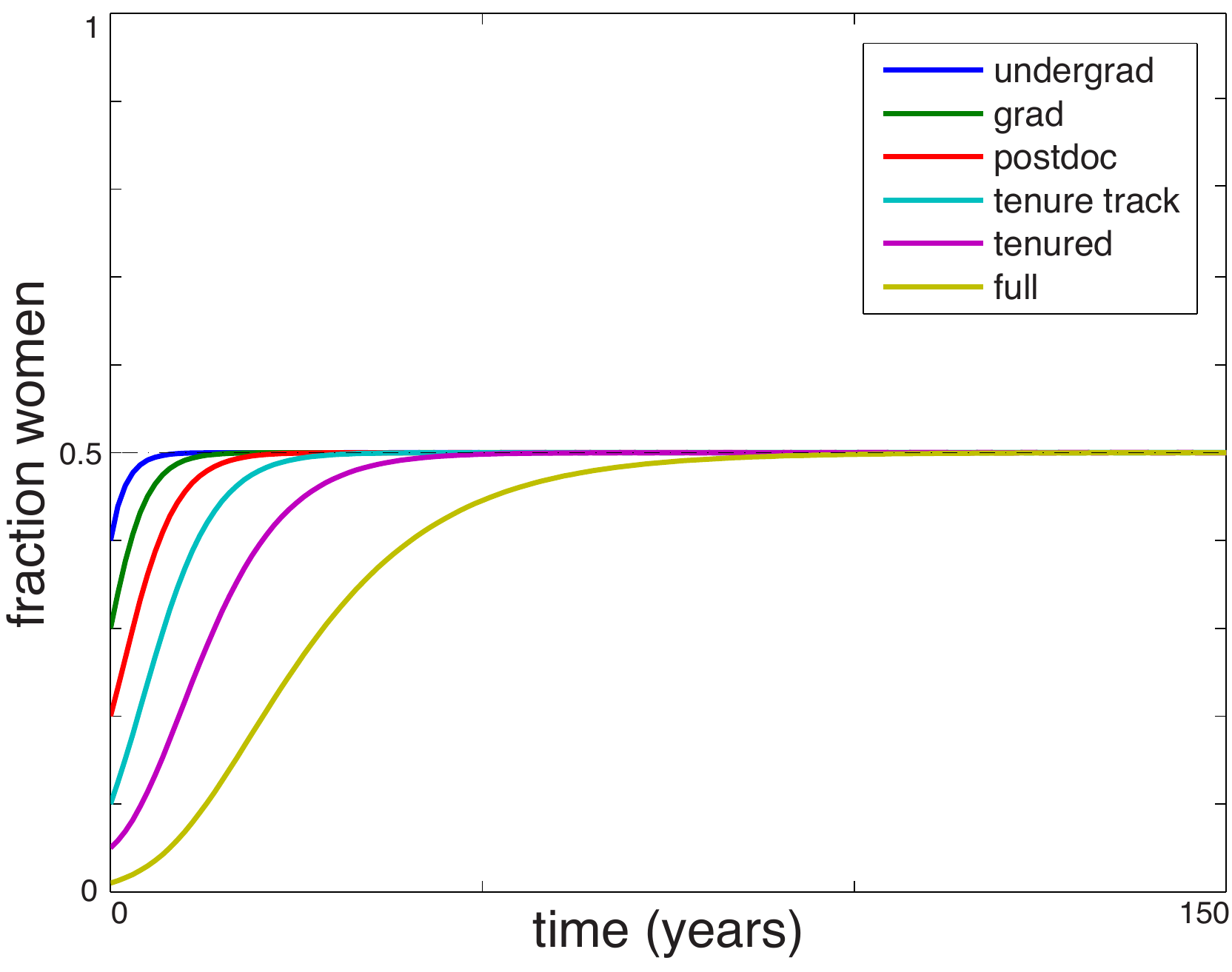}
  \end{center}
  \caption{Example of direct convergence to 50/50 gender split under the null model \eqref{eq:null}. In this example, we consider a hypothetical academic hierarchy with six levels, $\{R_j\} = \{1/4,1/5,1/6,1/7,1/9,1/15\}$, and $\{N_j\} = \{13,8,5,3,2,1\}$.}
  \label{fig:null}
\end{figure}

\subsection{Homophily-free model}
Now consider a model in which people do not use gender to decide whether to apply for a promotion (i.e., $\lambda = 0$), but employers are biased towards or against women (i.e., $b\ne \tfrac{1}{2}$). In this case, the model \eqref{eq:full} reduces to 
\begin{equation}
\frac{1}{R_j} \frac{\mathrm{d} x_j}{\mathrm{d} t} = (1+r_j) \frac{b x_{j-1}}{b x_{j-1} + (1-b) (1-x_{j-1})} - x_j -  r_j \frac{b x_{j}}{b x_{j} + (1-b) (1-x_{j})}  \,\,\,\,\,\,\, \text{for } 1 \le j \le L.
\label{eq:noseg}
 \end{equation}
As in the null model \eqref{eq:null}, the homophily-free model has a single, attracting fixed point. The presence of bias, however, pushes the steady-state gender fractionation away from gender parity. This effect is more extreme in higher levels than in lower ones. In particular, if the bias is against women ($b<\frac{1}{2}$),
\[
x^{*}_{L}<\ldots<x^{*}_{j}<\ldots<x^{*}_{1}<\frac{1}{2}.
\]
See the Appendix for details, and see Figure \ref{fig:noseg} for transient model behavior for a hypothetical academic hierarchy.

\begin{figure}[htb]
  \begin{center}
    \includegraphics[width=\textwidth]{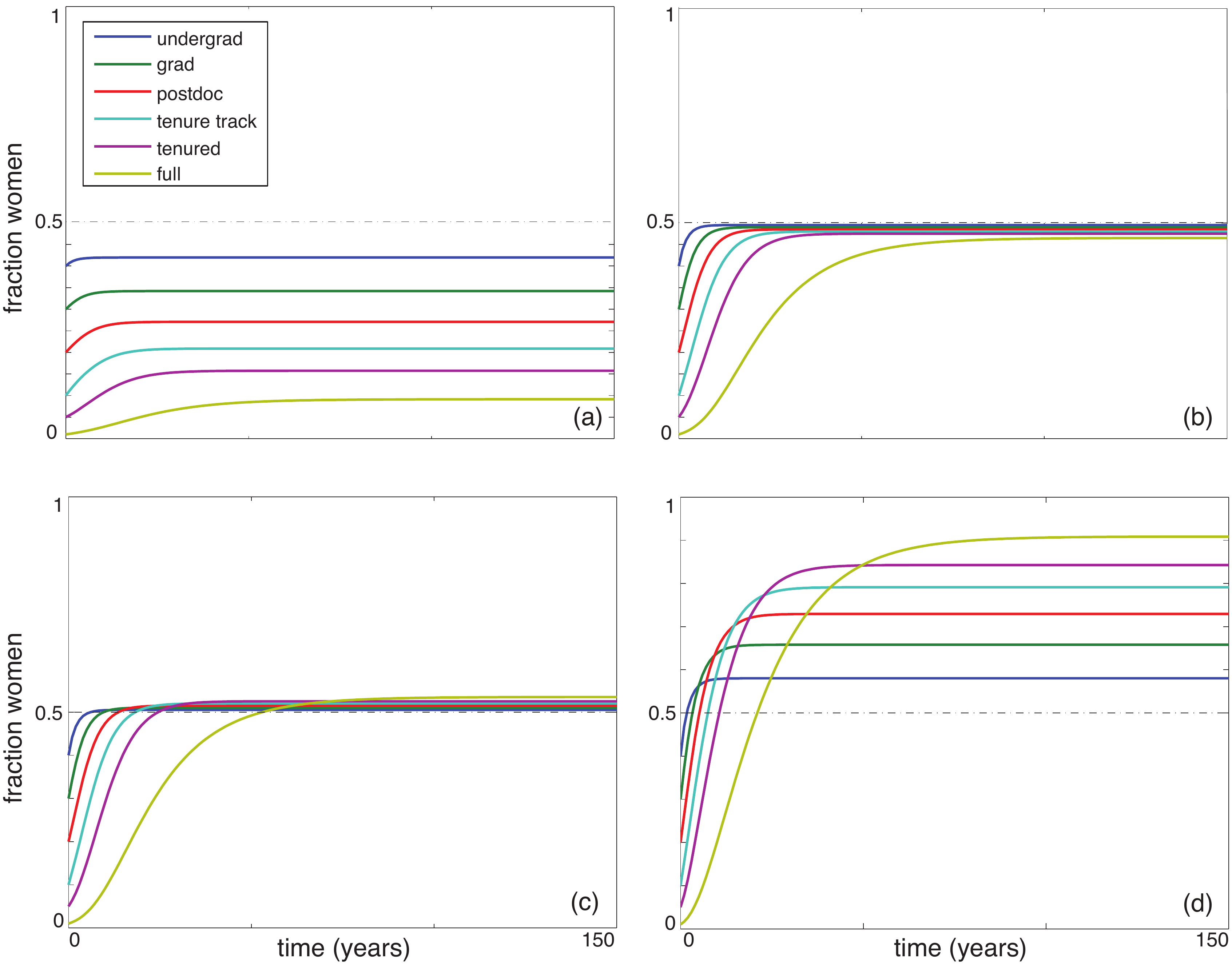}
  \end{center}
  \caption{Examples of transient behavior of a 6-level homophily-free model \eqref{eq:noseg}. \textbf{(a)} For strong bias against women ($b = 0.35$), all levels directly converge to male majority, with the strongest majority in the highest levels of leadership. \textbf{(b)} For weak bias against women ($b = 0.49$), the fraction of women in each level directly converges to a value near 50/50, though there are still more men in each level. \textbf{(c)} For weak bias favoring women ($b = 0.51$), the fraction of women in each level directly converges to a value near 50/50, though there are more women in each level. \textbf{(d)} For strong bias favoring women ($b = 0.65$), all levels directly converge to female majority, with the strongest majority in the highest levels of leadership.}
  \label{fig:noseg}
\end{figure}

\subsection{Bias-free model}
Consider an alternative model in which people self-segregate by gender, but employers are not biased towards or against women (i.e., $b=\tfrac{1}{2}$). Then model \eqref{eq:full} reduces to 
\begin{equation} 
\frac{1}{R_j}\frac{\mathrm{d} x_j}{\mathrm{d} t} = (1+r_j) f_0\left(x_j,x_{j-1}\right) - x_j -  r_j f_0\left(x_{j+1}, x_{j}\right) \,\,\,\,\,\,\, \text{for } 1\le j \le L.
\label{eq:biasfree}
\end{equation}
We observe three qualitatively different model behaviors for \eqref{eq:biasfree}: for mild homophilic tendencies, the system converges to gender parity; for moderate homophily, the fraction of women oscillates in all levels; and for strong homophily, the system converges to either male or female dominance depending on the initial state. The emergence of oscillations in such a system may not seem intuitively obvious. We explain the onset of oscillations in the Appendix.

Figure \ref{fig:nobias} shows the range of model behavior for a hypothetical academic hierarchy. See Figure \ref{fig:nobiasbif} for an example of a bifurcation diagram for the bias-free system. Although this diagram is representative of typical model behavior, the location of bifurcation points may shift as parameters vary.

\begin{figure}[htb]
  \begin{center}
    \includegraphics[width=\textwidth]{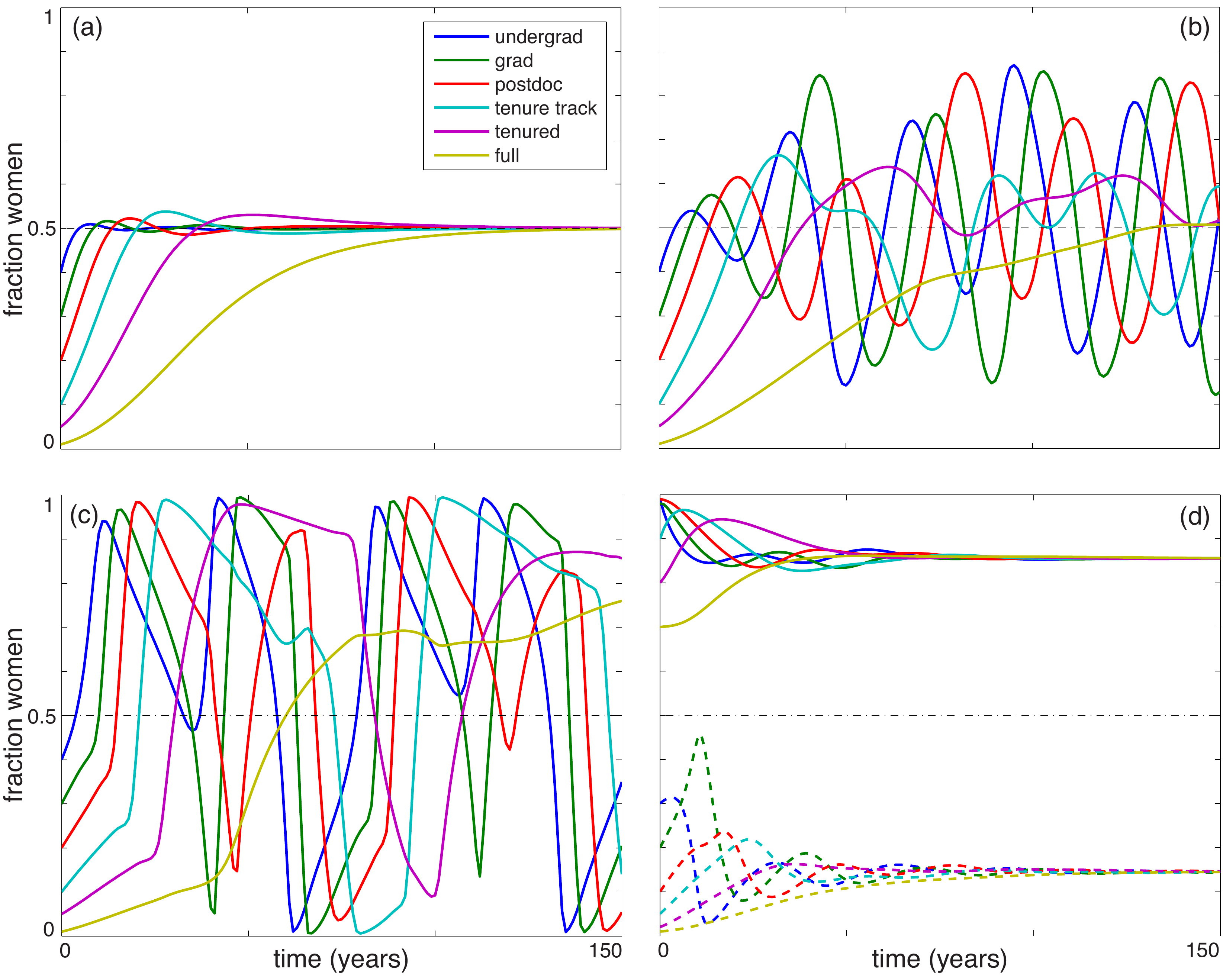}
  \end{center}
  \caption{Examples of transient behavior of a hypothetical academic 6-level bias-free model \eqref{eq:biasfree}. \textbf{(a)} For mild homophily ($\lambda = 2$), all levels converge to gender equity after oscillating above and below a 50/50 split.  \textbf{(b)} For stronger homophily ($\lambda = 3$), the fraction of women in each level oscillates about the 50/50 split without converging. \textbf{(c)} For yet stronger homophily ($\lambda = 4.5$), limit cycles appear to behave like those of a relaxation oscillator. \textbf{(d)} For strong homophily ($\lambda = 5$), each level equilibrates to nearly all women (solid lines) or nearly all men (dashed lines), depending on the initial condition. For all examples, $\{R_j\} = \{1/4,1/5,1/6,1/7,1/9,1/15\}$, and $\{N_j\} = \{13,8,5,3,2,1\}$.}
  \label{fig:nobias}
\end{figure}

\begin{figure}[htb]
  \begin{center}
    \includegraphics[width=\textwidth]{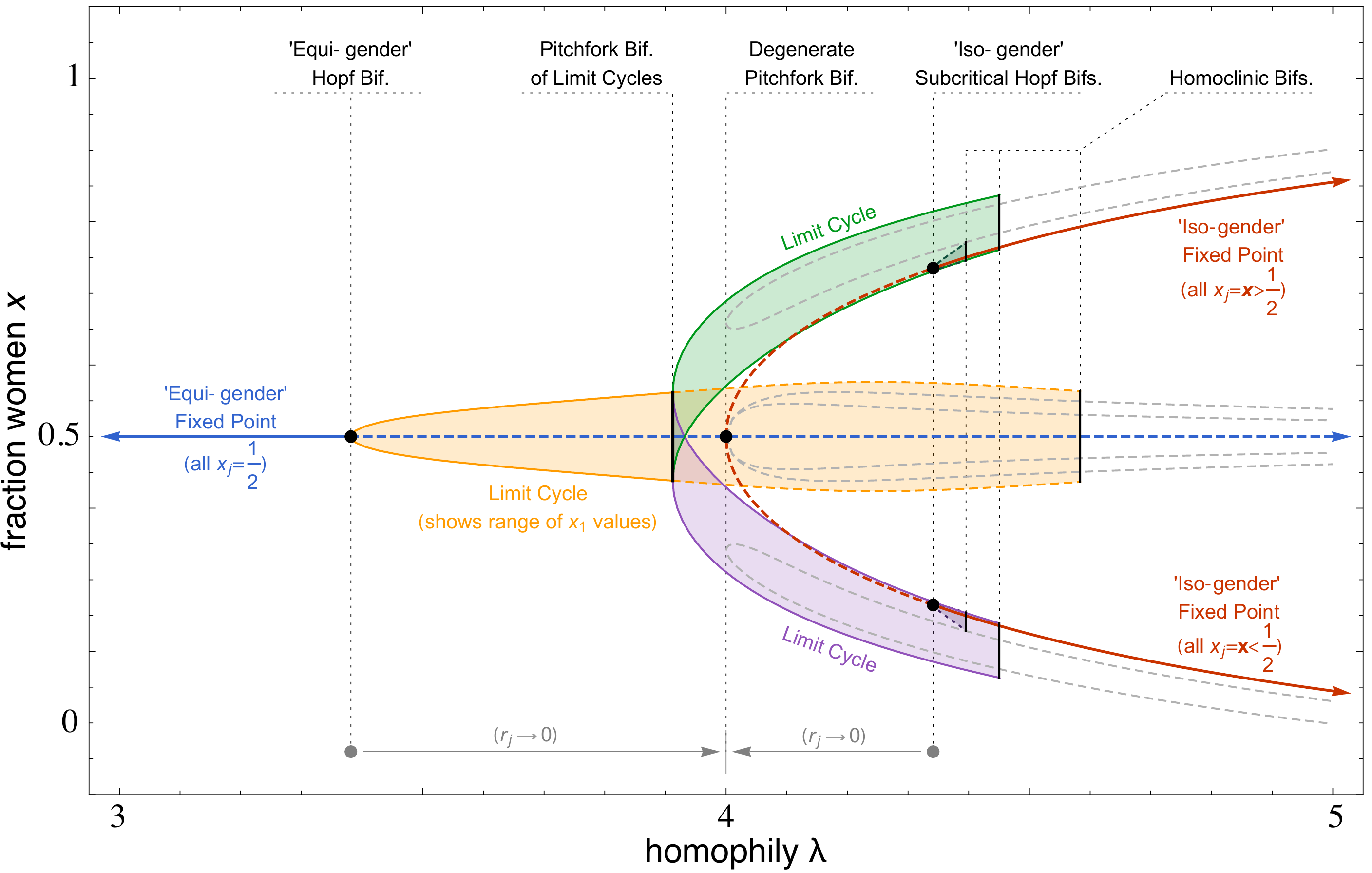}
  \end{center}
  \caption{Numerical bifurcation diagram for homophily parameter $\lambda$ in a 3-level bias-free system. Solid lines are stable equilibria/cycles, dashed lines are unstable equilibria/cycles, black dots are bifurcations of equilibria, and black lines are bifurcations of limit cycles. All limit cycles show the gender fractionation for the lowest level, $x_1$. Generated using AUTO\cite{doedel2007auto,ermentrout2002simulating} with $N_1=70, N_2=2, N_3=1, R_1=1/4, R_2=1/5, R_3=1/6$. Convergence to a degenerate pitchfork bifurcation at $\lambda=4$ as $r_j \to 0$ is shown in the Appendix.}
  \label{fig:nobiasbif}
\end{figure}

For the parameter values listed in the caption of Figure \ref{fig:nobias}, we see that as homophily increases from a small value, a supercritical Hopf bifurcation occurs, which initiates the onset of stable oscillations in all hierarchy levels. Although these oscillations are not identical, they have the same period at steady state, as suggested by the transient behavior in Figure \ref{fig:nobias}(b) and (c). 
At $\lambda\approx 3.9,$ the limit cycle in each hierarchy level undergoes a pitchfork bifurcation of limit cycles, after which no stable equilibria at or steady oscillations about gender parity occur.

At $\lambda= 4,$ a degenerate pitchfork bifurcation occurs for all parameter values. At this point, $2L+1$ equilibria, several of which are unstable, emanate from the pitchfork as determined by a center manifold reduction. In Figure \ref{fig:nobiasbif}, we focus on the equilibrium at gender parity and a pair of equilibria which eventually become stable, through subcritical Hopf bifurcations at $\lambda\approx 4.35.$ All limit cycles eventually end at homoclinic bifurcations: the periodic orbit spends more and more time near a saddle point (not shown) as the period diverges.

As $r_j \rightarrow 0$ for each level, the Hopf bifurcations converge at the pitchfork bifurcation. In that limit the pitchfork has a greater degeneracy, producing $3^{L}$ equilbria. Loosely speaking, hierarchies with small $r_j$ have very few people retiring relative to the number of people who would like to be promoted, making the hierarchies competitive. The limit $r_j \rightarrow 0$ is not realistic for any real-world hierarchy to our knowledge, but analysis near this limit aids numerical continuation; see the Appendix for details. 

\subsection{Model with homophily and bias}
\begin{figure}[htb]
  \begin{center}
    \includegraphics[width=\textwidth]{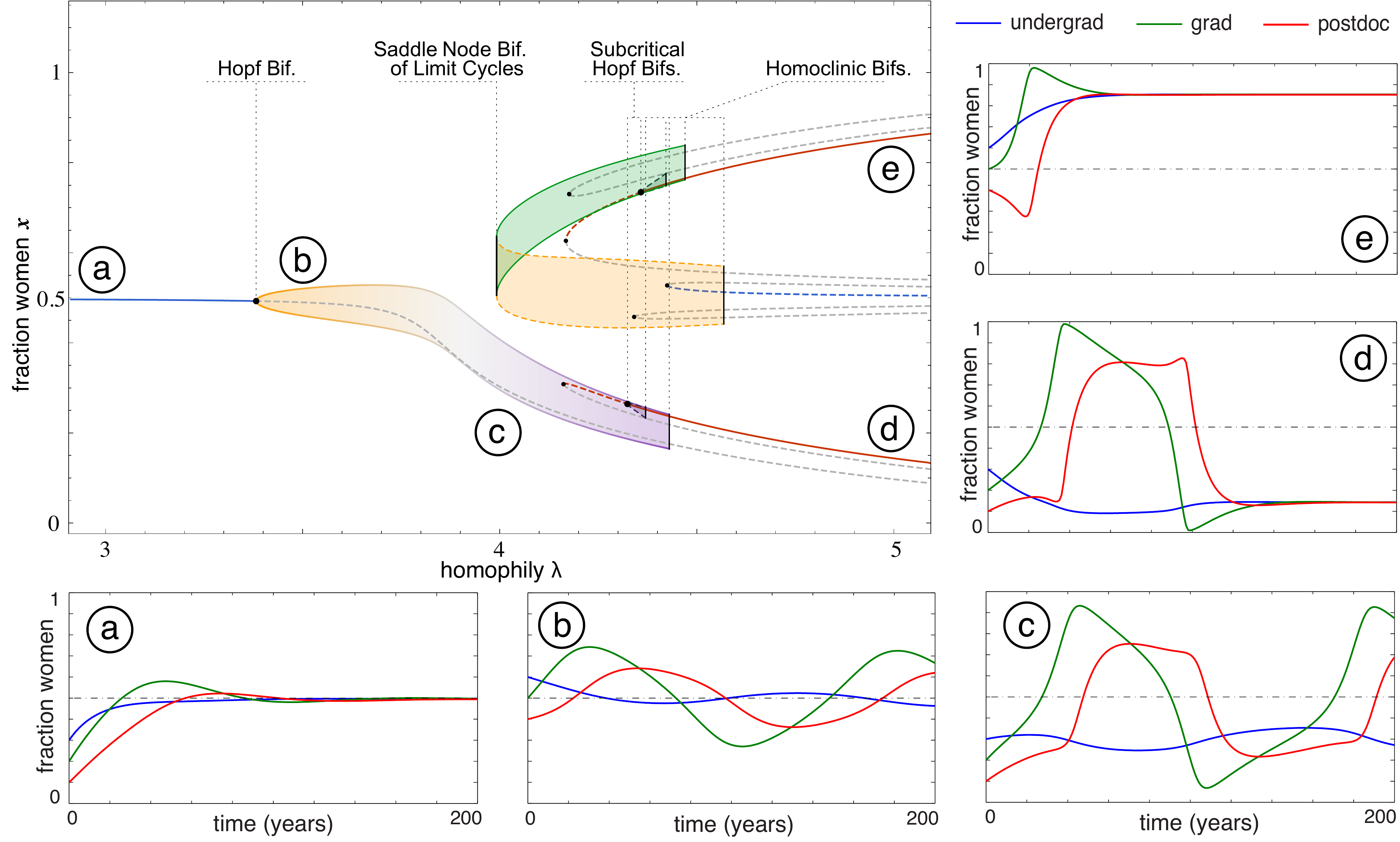}
  \end{center}
  \caption{Numerical bifurcation diagram for homophily parameter $\lambda$ in a 3-level system with slight bias against women ($b=0.499$). Solid lines are stable equilibria/cycles, dashed lines are unstable equilibria/cycles, black dots are bifurcations of equilibria, and black lines are bifurcations of limit cycles. All curves show the gender fractionation for the lowest level, $x_1$. Generated using AUTO\cite{doedel2007auto,ermentrout2002simulating} with $N_1=70, N_2=2, N_3=1, R_1=1/4, R_2=1/5, R_3=1/6$. Examples of transient behavior for several positions within the bifurcation diagram are on the margins: (a) $\lambda = 3$, (b) $\lambda = 3.5$, (c) $\lambda = 4$, (d) $\lambda = 5$ with lower initial condition, and (e) $\lambda = 5$ with higher initial condition.}
  \label{fig:bif2}
\end{figure}

Finally, we explore the full model \eqref{eq:full} with bias $b\ne \tfrac{1}{2}$ and homophily $\lambda\ne0$. The long-term dynamics are similar to those of the bias-free model \eqref{eq:biasfree}. For small homophily, regardless of initial state, the hierarchy tends towards a `biased' fractionation profile. For large homophily, the gender fraction polarizes, with bistable equilibria at both large and small fractions of women at each level. Figure \ref{fig:bif2}(a),(d),(e) show examples of transient behavior at these high and low homophily values. 

Figure \ref{fig:bif2} shows a slight perturbation of the system from the bias-free case, highlighting the degeneracy of the pitchfork bifurcation in Figure \ref{fig:nobiasbif}; branch colors correspond with the colors of related branches in Figure \ref{fig:nobiasbif}. Generically, for moderate levels of homophily, the limit cycles that emanate from the bifurcation `bend' in the direction of bias (e.g., for $b<\frac{1}{2},$ toward fewer women in each hierarchy level) as homophily increases through the supercritical Hopf bifurcation. The degenerate pitchfork bifurcation unfolds into several saddle-node bifurcations and a continuous fixed point curve. Similarly, the pitchfork bifurcation of limit cycles unfolds into a saddle-node bifurcation of limit cycles and a continuous limit cycle curve. As in Figure \ref{fig:nobiasbif}, all limit cycles end in homoclinic bifurcations.

For lower values of bias $b$, the Hopf bifurcation from the equigender fixed point shifts along the branch of equilibria it emanates from, corresponding to a decrease in $x_1$ and an increase in homophily. At the same time, the length of the limit cycle branches emanating from the Hopf point decreases, and the Hopf point is eliminated in a Takens-Bogdanov bifurcation. For stronger bias ($b\approx 0.45$), long-term behavior manifests solely as equilibria, which includes the possibility of decaying oscillations. Limit cycles are no longer possible. See the Appendix for the co-dimension 2 bifurcation diagram, where both bias $b$ and homophily $\lambda$ are varied.

\section{Model validation}
With this simple model, we aim to extract useful information from real-world hierarchies without claiming to fully explain their dynamics. For instance, we wish to predict when (or if) fields will reach gender parity, what sociological or psychological factors may be the main drivers of gender fractionation dynamics, and what interventions may help various fields reach gender parity more quickly. 
\subsection{Data}
We collect time series data on the fraction of women in each level of many professional hierarchies \cite{nsf2017ncses,nsf2017nss,ies2017nces,acs2015cen,nasem2013seeking,apa2017sd,kff2017med,aamc2016med,abim2018med,brotherton2017graduate,brotherton2016graduate,brotherton2015graduate,topaz2016gender,lab_2017,asne2015,blount1998destined,dana2006women,cat2018www,aba2013law,nalp2018law,nawl2017law,nawl2018rut,pew2018gov,msu2017cihws,uscb2013nurse,hrsa2008nurse,lao2016orch}.
Although most studies of this nature have focused on academia \cite{holman2018gender,shaw2012leaks}, the generality of our model allows us to examine a larger variety of hierarchies: medicine, law, politics, business, education, journalism, entertainment, and fine arts/music. Of the 23 hierarchy datasets we assembled, 16 are sufficiently comprehensive to attempt model fitting. Each dataset comprises the following components:
\begin{itemize}
\item A hierarchy structure (e.g., undergraduate $\to$ graduate $\to$ postdoctoral $\to$ assistant professor $\to$ associate professor $\to$ professor, in a typical academic hierarchy). In the real world, the hierarchical structure is not perfectly rigid, but we take the structure to be the `typical' route through the ranks. This structure determines the hierarchy size $L$ and the ordering of levels in our model \eqref{eq:full}.
\item The fraction of each level of the hierarchy that are women over time. We include datasets with at least a decade's worth of continuous yearly data for all levels. If there are missing years, we use linear interpolation to fill the gaps. Some datasets were available in a table, but others were extracted from graphical representations using WebPlotDigitizer \cite{rohatgi2018webplotdigitizer}. This determines the exact $x_j(t)$ for a range of discrete times.
\item The approximate relative sizes of each level. Although fields may grow (e.g., medicine) or shrink (e.g., journalism) over time, we find that the relative level sizes generally stay approximately the same. Where data on the relative level sizes were not available, we made educated guesses. This information estimates $N_j$ in our model if we normalize the top level to $1$ ($N_L = 1$).
\item The approximate yearly `leave' or `retirement' rates for each level. These statistics are not available for any hierarchies, to our knowledge. We made educated guesses for these parameters based on the expected amount of time spent in each level. For instance, the vast majority of undergraduate degrees are completed in approximately four years, and relatively few graduates continue on to doctoral study. Therefore, our initial estimate for the undergraduate leave rate is $0.25$ (i.e., approximately a quarter of undergraduates leave college each year without moving up the academic hierarchy). We take these proxies to exit rates as estimates for $R_j$ in our model.
\end{itemize}
All compiled data, including datasets not sufficient for model fitting, are available at Northwestern's ARCH repository: \url{https://doi.org/10.21985/N2QF28}.

\subsection{Model fitting}
We wish to fit the model to each dataset in order to quantify the degree of bias and homophily in each field; with this information, we may predict the long-term fraction of women in each level of the hierarchies without any intervention, and we can suggest targeted interventions to reach gender parity more quickly. Theoretically, distinguishing between bias and homophily in the data should be straightforward because the qualitative effects of each parameter are different. Bias is the only parameter that independently `separates' levels (i.e., bias causes the female fractionation $x_j$ to differ among levels), while homophily is the only parameter that independently causes oscillations.

There are many possible ways to fit the model to each dataset. One qualitative way to measure the degree of bias and homophily in each dataset is to look for separation between levels and indications of oscillations. Roughy speaking, datasets with strong bias either towards or against women will have large changes in the proportion of women as one ascends the hierarchy (e.g., see Figure \ref{fig:noseg}(a),(d)). 

On the other hand, datasets with weak bias and moderate homophily will show signs of oscillations in each level (e.g., see Figure \ref{fig:nobias}(b),(c)), although real datasets may not include enough time points to resolve a full period of the oscillations. Datasets with weak bias and strong homophily will appear male- or female-dominated without much separation between levels (e.g., see Figure \ref{fig:nobias}(d)). If both bias and homophily are strong, then the impact of each phenomenon will be difficult to deduce visually (see Appendix for phase diagram), and quantitative methods will be needed. 

As a quantitative attempt at fitting, we perform a global minimization of error between the model and data. We first find a best fit of the model to each dataset by minimizing the sum of squared error between the model gender fractionation $\hat{x}_j$ and the data $x_j$ over time using the Nelder-Mead minimization algorithm \cite{nelder1965simplex}. The fitting parameters are $b, \lambda, R_j, N_j$ and the initial conditions. We include $R_j$ and $N_j$ as fitting parameters because we do not have exact values for these parameters, but we heuristically verify that the model fit does not select values far from our initial guesses. The initial condition is a fitting parameter to ensure the first data point does not contribute more weight to the fitting process than the subsequent data points in the time series.  

We seed the Nelder-Mead algorithm with 20 initial guesses for the fitting parameters $b$ and $\lambda$, selected uniformly from $b \in (0.2,0.8)$ and $\lambda \in (1.5, 6.5)$. All other parameter guesses are taken to be our best estimates from available data. After finding the best fit parameters $(\tilde{b}, \tilde{\lambda})$ from among the 20 seeded searches, we run a second search in the parameter space near the best fit. In this next step, we seed the Nelder-Mead algorithm with 10 new initial guesses for $b$ and $\lambda$, selected from normal distributions $b \in \mathcal{N} (\tilde{b}, 0.05)$ and $\lambda \in \mathcal{N} (\tilde{\lambda}, 0.1)$. We take, as our final fit, the best fit parameters after this second search. See Appendix for a visual representation of this algorithm.

\begin{figure}[htb]
  \begin{center}
   \includegraphics[width=\textwidth]{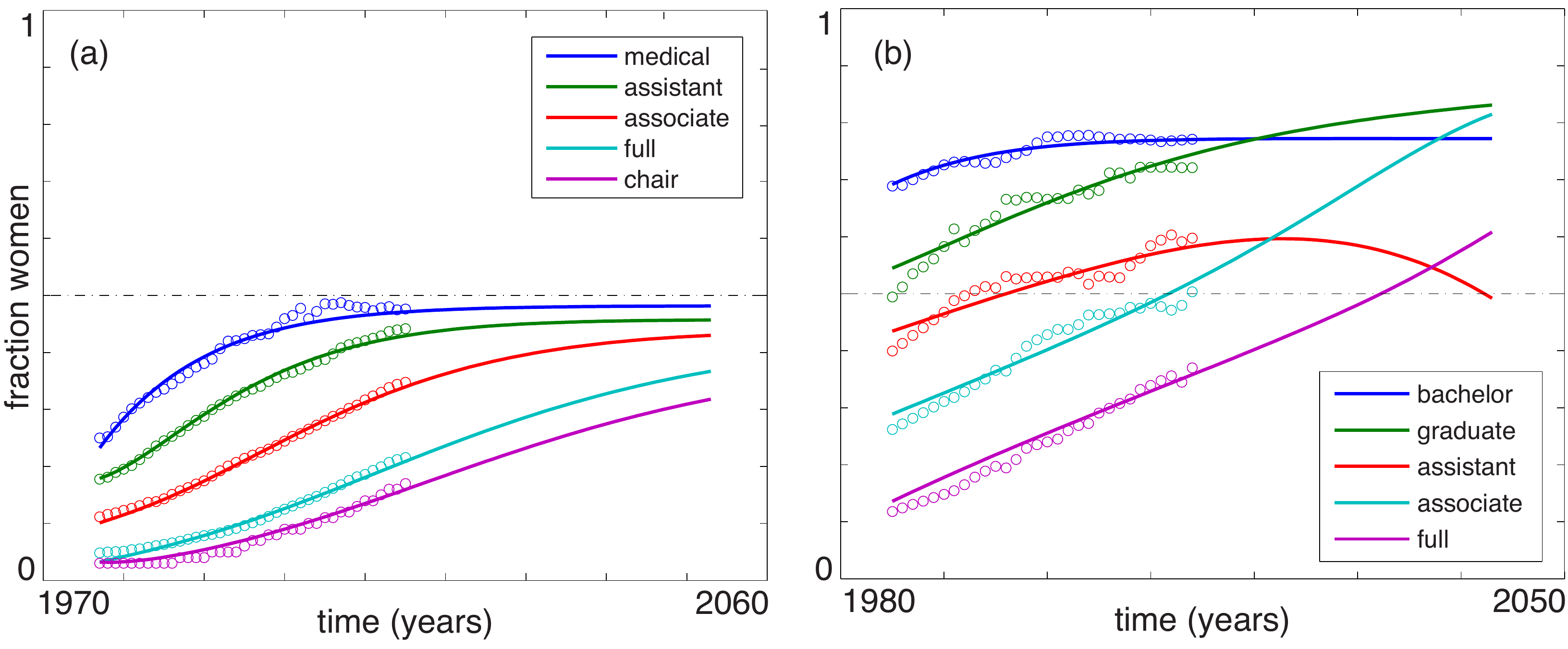}
  \end{center}
  \caption{Model fit to data from (a) clinical academic medicine\cite{kff2017med,abim2018med,aamc2016med} and (b) academic psychology\cite{nsf2017nss,ies2017nces,apa2017sd}.}
  \label{fig:fit1}
\end{figure}

We present the best fits from two representative hierarchies in Figure \ref{fig:fit1}. Best fit parameters $\hat{b}$ and $\hat{\lambda}$ from all datasets are shown in Figure \ref{fig:bvh}. See Appendix for fit parameters and additional model predictions for all datasets.

\begin{figure}[htb]
  \begin{center}
    \includegraphics[width=\textwidth]{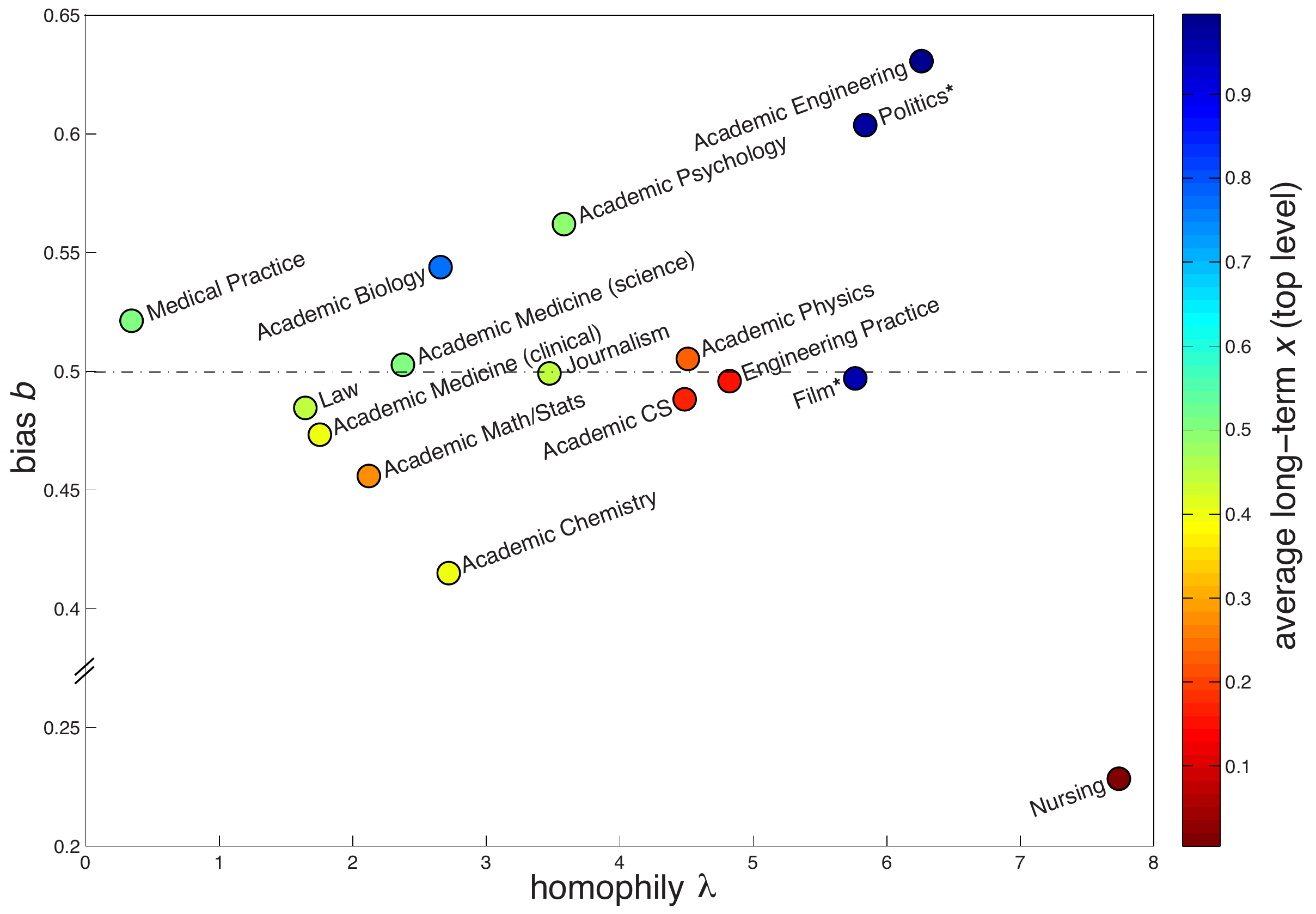}
  \end{center}
  \caption{Bias and homophily best fit parameters for each hierarchy. Colors indicate the predicted long-term (equilibrium) female fractionation in the highest level of leadership; if the hierarchy is not predicted to reach equilibrium, then a time average over the limit cycle was taken.
  *May not be a strict hierarchy: although producers hire directors, producers do not typically `promote' directors to producer positions. Likewise for politics.}
  \label{fig:bvh}
\end{figure}

To address the concern of possible overfitting, we verify that the ratio of data points to fitting parameters is large. For each dataset, there are $3 L+1$ fitting parameters and $L T$ data points, where $L$ is the number of levels and $T$ is the number of years in the dataset. The datasets with the fewest number of levels available and fewest years available should prompt greatest concern regarding overfitting. Among our datasets, the smallest ratio of parameters to data points was for the journalism hierarchy, which had 51 data points and 10 parameters. The typical ratio of data points to parameters was about 10:1.

Because our parameter search algorithm is not guaranteed to find the absolute minimum error between the model and data, we verify that our model results are not excessively sensitive to changes in our fitting procedure. To illustrate, we seed our algorithm's random number generator with ten different seeds and verify that the variation in predicted average gender fractionation is small. We select the two most concerning datasets for this computationally intensive test: (1) journalism, due to its risk for overfitting, and (2) academic engineering, due to its unpredictable fitting results during early tests (see Appendix).

\section{Discussion}
The presented model vastly simplifies the process by which people choose to advance their careers, yet we may exploit the model to extract useful predictions and suggestions for interventions to reach gender parity. By fitting the model to data from over a dozen professional hierarchies, we may predict the time required to reach gender parity if there are no cultural or policy shifts within the fields. Unlike the model by Holman et al. \cite{holman2018gender}, we predict that many fields may never reach gender parity without intervention (see Appendix). For instance, fields that indicate especially strong homophily (e.g., engineering and nursing) are expected to become male- or female-dominated. Fields with apparently strong bias against women (e.g., academic chemistry, math, and computer science) are predicted to never reach sustained gender parity, at least in the highest levels of leadership. 

Fields with bias near $1/2$ and weak homophily (e.g., medicine and law) are predicted to eventually reach gender parity as fast as inertia allows, as modeled by Shaw and Stanton \cite{shaw2012leaks}. Effective affirmative action programs could artificially speed the process, but resources may be better spent in fields where gender parity is not inevitable. One benefit of our modeling approach is that we can extract the relative impact of two major decision-makers in a professional hierarchy: those who apply for promotion and those who grant promotion. For fields with strong bias against women ($b<1/2$), the decision-makers that should be targeted are hiring committees. For instance, hiring committees could be trained in unconscious bias, or policies could mandate that the number of promotions offered to women match the applicant pool. For fields with strong homophily, the decision-makers that should be targeted are women eligible for promotion. Knowing that fewer women than are eligible are applying for promotion in male-dominated fields, hiring committees could actively recruit women to apply for promotion or make the underrepresented gender more visible within the field.

\subsection{Limitations}
Of course, the predictions and interventions suggested by this simple model are subject to limitations. We assume that hierarchical structures remain constant over time, but this is not always the case. For instance, some fields that now require a college degree were once accessible to those with a high school education. We also assume that individuals must pass through each level linearly, but many academic fields may or may not include a postdoc, and political or business leaders may come from outside their field entirely.

To avoid overfitting, we assume that bias and homophily are constant both across time and across the hierarchy structure. Naturally, the cultures and policies that shape these sociological properties are not constant; perhaps bias against women has diminished over time, but maybe bias is stronger at higher levels of leadership. Also, gender may be more salient to a young person deciding on a major than on an associate professor up for promotion. Therefore, we think of the fitting parameters $\hat{b}$ and $\hat{\lambda}$ as an average bias and homophily over time and the hierarchy structure. 

Finally, we have ignored the different decisions that men and women may make. Our model assumes that men and women on hiring committees are equally biased against a certain gender, that gender is equally salient to men and women, and that men and women are equally qualified for advancement. A more sophisticated model may break the symmetry between men and women. 

\subsection{Future steps}
Allowing bias and homophily to change over time and across the hierarchy structure is a natural model extension. In addition to making the model more realistic, it would also permit interventions to be incorporated directly into the model. If the effect of an intervention is to change bias and/or homophily, then the model could serve as the basis of a control problem to find an optimal time-dependent intervention.

Due to the generality of the model, it could also be extended to study the progression of underrepresented minorities through professional hierarchies. A few complications are introduced in this case: our model assumes that the gender distribution of the general population is constant in both space and time, but for racial minorities this is not true. Also, data collection may prove to be more complicated due to the evolving and sometimes overlapping definitions of various racial and ethnic groups.


Finally, the model could be generalized to include a spectrum of gender identities, income levels, or socioeconomic privilege. Two major challenges are introduced with this model extension. First, the current system of $L$ ordinary differential equations may become a system of $L$ partial integro-differential equations, which will make model analysis more difficult. Second, data required to validate such a model will be more challenging to obtain.

\section{Conclusion}
We have developed a simple model of the progression of people through professional hierarchies, like academia, medicine, and business. The model assumes that gender is a salient factor in both the decision to apply for promotion and the decision to grant promotion, but that men and women do not make fundamentally different decisions. Unlike previous models of the phenomenon, our model predicts that gender parity is not inevitable in many fields. Without intervention, a few fields may even become male- or female-dominated in the long term.

By fitting our model to available data, we extract the relative impact of the major decision-makers in the progression of women through 16 professional hierarchies. In some fields, like academic chemistry, bias of promotion and hiring committees may be the dominant reason that women are poorly represented. In other fields, like engineering, women not applying for promotion may be the dominant reason for the so-called leaky pipeline. With this information, we may suggest effective interventions to reach gender parity.

\section{Acknowledgements}
The authors wish to thank Danny Abrams, Yuxin Chen, Stephanie Ger, Joseph Johnson, and Rebecca Menssen for valuable conversations during the model development stage. Thanks are also due to Elizabeth Field, Alan Zhou, and the Illinois Geometry Lab for contributions to and support of model analysis and data collection. The authors additionally thank Chad Topaz for offering comments that greatly improved the manuscript.

The authors also wish to thank Jo\~{a}o Moreira (Amaral Lab, Northwestern University), Peter Buerhaus and Dave Auerbach (Center for Interdisciplinary Health Workforce Studies, Montana State University), Roxanna Edwards (Bureau of Labor Statistics), and Karen Stamm (Center for Workforce Studies, American Psychological Association) for sharing unpublished data.

This work was funded in part by National Science Foundation Graduate Research Fellowship DGE-1324585 and Mathways grant DMS-1449269 (SMC), Royal E.~Cabell Terminal Year Fellowship (KH, EEA), and National Science Foundation Research Training Grant DMS-1547394 (AJK). The funders had no role in study design, data collection and analysis, decision to publish, or preparation of the manuscript.

\section{Data availability}  
All data (Excel .xlsx file) and software (Matlab .m files and XPPAUT .ode files) are publicly available from the Northwestern ARCH repository (DOI:10.21985/N2QF28) at \url{https://doi.org/10.21985/N2QF28}.

\section{Competing interests}
The authors declare no competing interests.

\bibliographystyle{ieeetr}
\bibliography{prohierarchyLibrary}

%


\newpage
\section{Appendix}
\subsection{Model Development}
To further illustrate the model, we derive the null model (no bias and no homophily) with a small number of levels. Consider a four level hierarchy (Figure \ref{fig:null}). The top level has $x_4$ fraction women, has $N_4$ total people, and therefore has $N_4 x_4$ women and $N_4(1-x_4)$ men. People in the top level retire at a rate $R_4$, so $R_4 N_4 x_4$ women and $R_4 N_4 (1-x_4)$ men leave the top level per time unit. 

\begin{figure}[htb]
  \begin{center}
    \includegraphics[width=\textwidth]{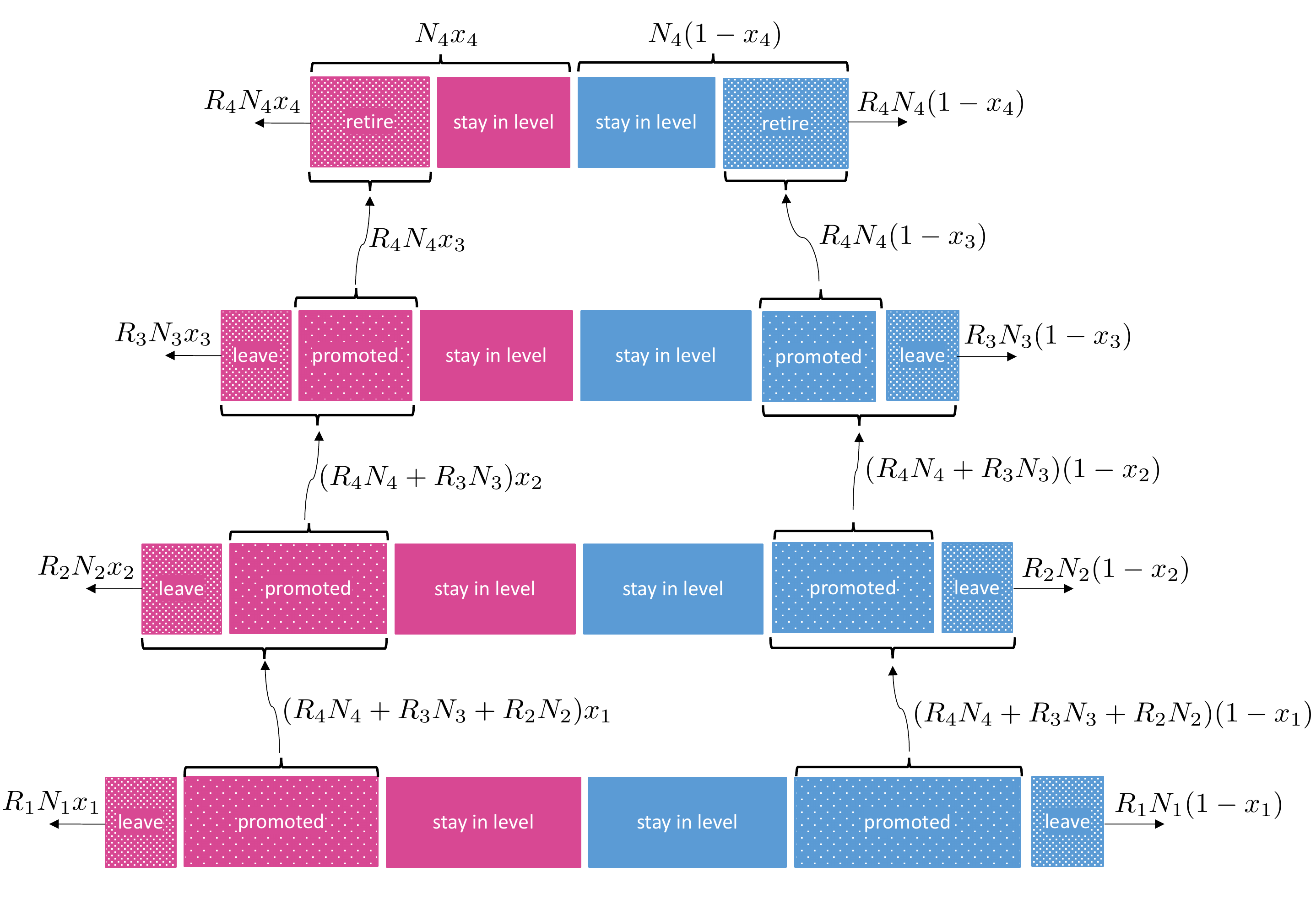}
  \end{center}
  \caption{Illustration of a four-level hierarchy without bias and homophily. Each horizontal bar is a level with $N_j$ people in it, $N_j x_j$ of whom are women. Pink bars represent the number of women in the level, and blue bars represent the number of men in the level. Dotted bars represent the number of people who are exiting the level, either via promotion or by leaving the hierarchy.}
  \label{fig:null}
\end{figure}

The $R_4 N_4$ people that retire from the top level must be replaced by people from the third level. Without any bias or homophily, the number of women promoted into the top level is proportional to the number of women in the third level. Therefore, $R_4 N_4 x_3$ of those promoted are women and $R_4 N_4 (1-x_3)$ are men. In addition to the $R_4 N_4$ people who are promoted out of the third level, another $R_3 N_3$ people leave the hierarchy from the third level without being promoted. Of those who leave the third level, $R_3 N_3 x_3$ are women, and $R_3 N_3 (1-x_3)$ are men.

A total of $R_4 N_4+R_3 N_3$ vacancies are created in the third level by those who are promoted and those who leave the hierarchy. These positions are filled by those in the second level. Without any bias or homophily, the number of women promoted into the third level is proportional to the number of women in the second level. Therefore, $(R_4 N_4 +R_3 N_3) x_2$ of those promoted are women. In addition to those promoted, another $R_2 N_2$ people, $R_2 N_2 x_2$ of whom are women, leave the hierarchy from the second level without being promoted. 

Thus, a total of $R_4 N_4+R_3 N_3 + R_2 N_2$ vacancies are created in the second level by those who are promoted and those who leave the hierarchy. Without any bias or homophily, $(R_4 N_4+R_3 N_3 + R_2 N_2) x_1$ of those positions are filled by women. This illustrates that demographic changes in lower levels depend on the total number of people leaving the hierarchy in higher levels.

Adding up the total change in the number of women in each level leads to the model \eqref{eq:derive} with $L=4$ and $f(x_j,x_{j-1};b) = x_{j-1}$.

\subsection{Model Behavior}
\begin{figure}[htb]
  \begin{center}
    \includegraphics[width=0.9\textwidth]{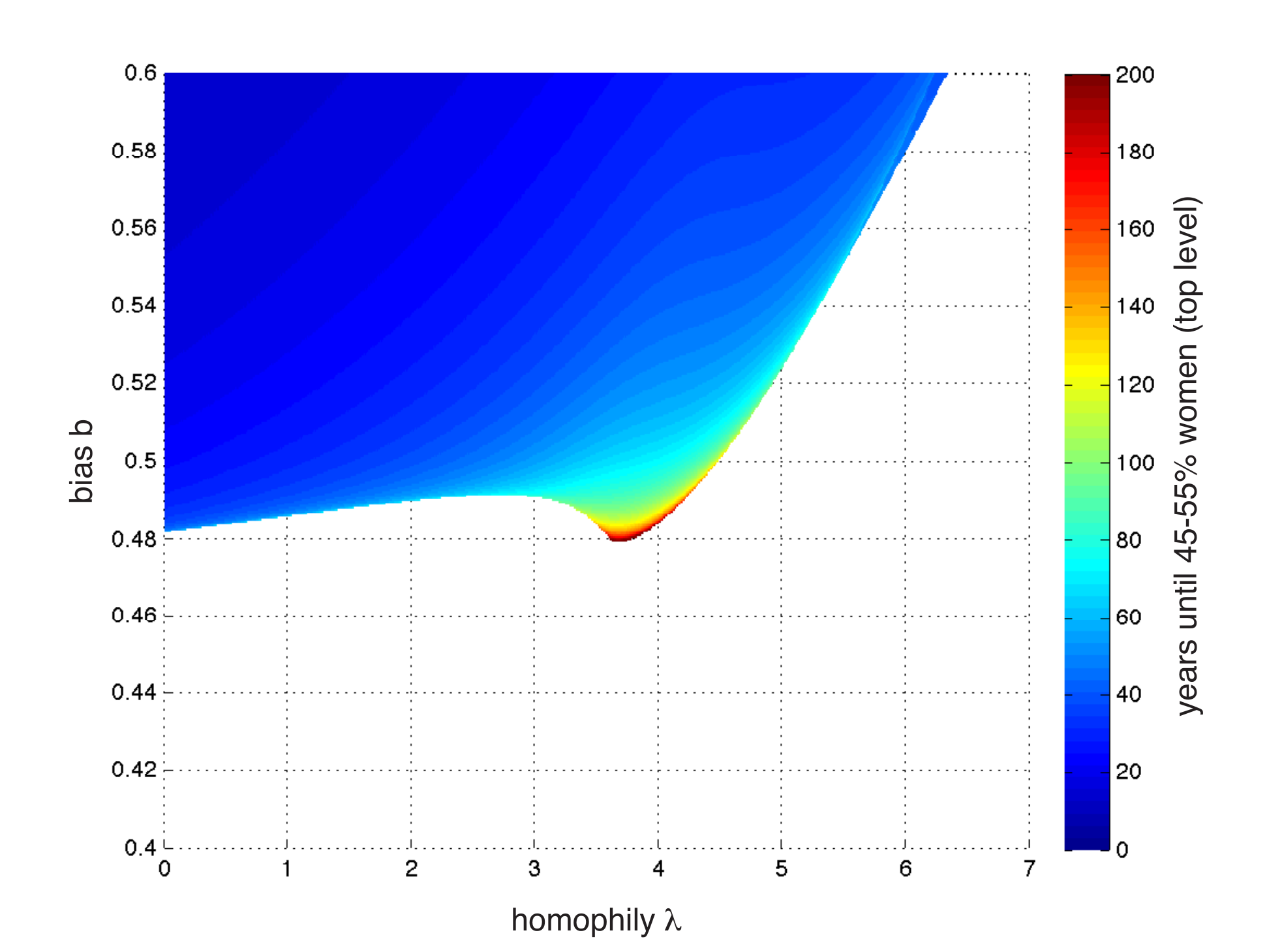}
  \end{center}
  \caption{Time that the top level of a 3-level hierarchy takes to reach within $5\%$ of parity (i.e., when $x_3\geq 0.45$) for a range of bias and homophily values. The bias parameter $b$ and homophily parameter $\lambda$ are varied, while the parameters $N_1=70$, $N_2=2$, $N_3=1$, $R_1=1/4$, $R_2=1/5$, $R_3=1/6$, and initial condition $(x_1(0),x_2(0), x_3(0))=(0,0,0)$ are held constant across all simulations. The white region corresponds to values of $b$ and $\lambda$ where the top level never comes within $5\%$ of parity.}
  \label{fig:topTime}
\end{figure}

\begin{figure}[htb]
  \begin{center}
    \includegraphics[width=0.9\textwidth]{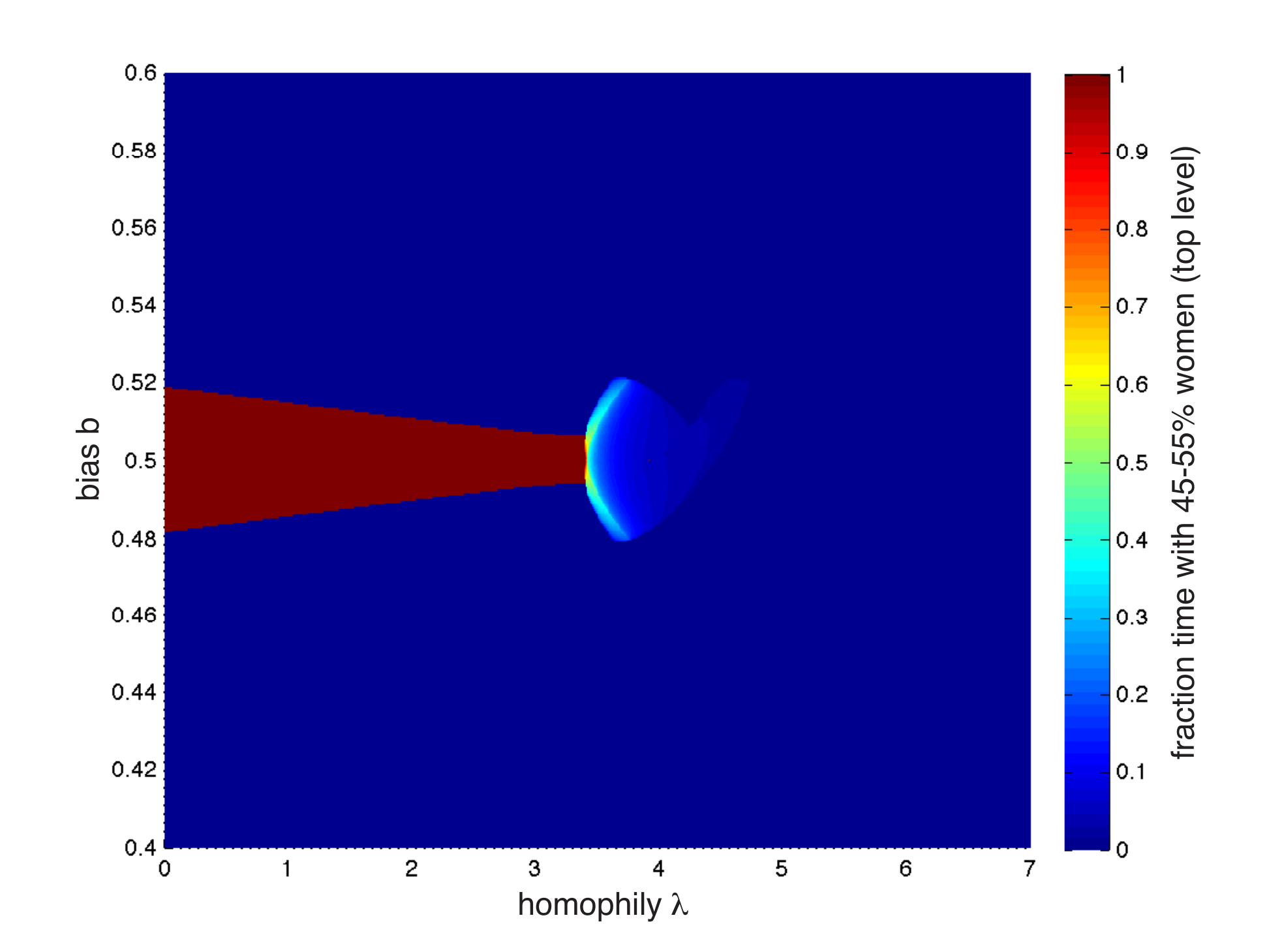}
  \end{center}
  \caption{Fraction of time that the top level of a 3-level system spends within $5\%$ of parity after the transient period (i.e., the fraction of time that $0.45\leq x_3\leq 0.55$) for a range of bias and homophily values. The bias parameter $b$ and homophily parameter $\lambda$ are varied, while the parameters $N_1=70$, $N_2=2$, $N_3=1$, $R_1=1/4$, $R_2=1/5$, $R_3=1/6$, and initial condition $(x_1(0),x_2(0), x_3(0))=(0,0,0)$ are held constant across all simulations. The dark blue region corresponds to values of $b$ and $\lambda$ where the top level reaches a steady state outside gender parity. The dark red region corresponds to parameter values where the top level reaches a steady state within a band of gender parity. All other colors indicate regions where the fraction of women in the top level oscillates.}
  \label{fig:topBand}
\end{figure}

Figures \ref{fig:topTime} and \ref{fig:topBand} display examples of stable model predictions for a wide range of parameter values. Figure \ref{fig:topTime} shows the time required for the top level of the hierarchy to get within 5\% of gender parity, a metric used in the analysis by Holman et al. \cite{holman2018gender}. Figure \ref{fig:topBand} shows the fraction of time the top level spends in a 5\% band about gender parity. 

\subsection{Model Analysis}
For efficient analysis, we define
\begin{align*}
Q(u-v) & \equiv \frac{P(1-u,1-v)}{P(u,v)}=\mathrm{e}^{-\lambda (u-v)},\\
Q_{0}(v) & \equiv \frac{1-v}{v}, \\
B & \equiv Q_{0}(b) = \frac{1-b}{b},
\end{align*}
where $P$ is the probability that a person applies for promotion. Note that $Q\left(0\right)=Q_{0}\left(\frac{1}{2}\right)=1$ and that $Q$  and $Q_{0}$ are monotonically decreasing. We can rewrite $f$ as
\begin{equation*}
f(x_{j},x_{j-1};B) = \frac{1}{1 + B Q_{0}(x_{j-1}) Q(x_{j}-x_{j-1})}\equiv f_{j}.
\end{equation*}
For the analyses in the following sections, it is useful to note that $\frac{\partial f_{j}}{\partial x_{j}} = \lambda f_{j} \left(1-f_{j}\right)$, that $\frac{\partial f_{j}}{\partial x_{j-1}} = \left(\frac{1}{x_{j-1} \left(1-x_{j-1}\right)} - \lambda\right)f_{j} \left(1-f_{j}\right)$, and that the evolution equations for $x_{j},\,j\in\{1,2,\ldots,L\}$ can be written as
\begin{align}
\frac{dx_{j}}{dt}=R_{j}\left[(1+r_{j})f_{j}-x_{j}-r_{j}f_{j+1}\right],
\end{align}
with $x_{0}=\frac{1}{2}$ and $r_{L}=0$.

\subsubsection{Stability of the Fixed Point in the Null and Homophily-Free Models}
In the special cases of the homophily-free ($\lambda=0$) and null ($\lambda=0, b=1/2$) models, each level's dynamics depends only on the fraction of women in that level and the level below. The Jacobian is therefore lower triangular, and the eigenvalues appear along the diagonal,
\begin{eqnarray*}
\lambda_{j} & = & -R_{j}\frac{\partial}{\partial x_{j}}\left(x_j +  r_j f(\cdot,x_{j};B)\right)\rvert_{x_{j}=x^{*}_{j}} \\
 & = & -R_{j}\left(1+r_{j} \frac{f^{*}_{j+1} \left(1-f^{*}_{j+1}\right)}{x^{*}_{j} \left(1-x^{*}_{j}\right)}\right),
\end{eqnarray*}
where the placement of $\cdot$ reflects the fact that $f$ does not depend on the first argument in this case, and $f^{*}_{j+1}$ denotes the value of $f_{j+1}$ for $x_{j}=x^{*}_{j}$. Since these values are all strictly negative, the single fixed point that appears in the absence of homophily is attracting.

\subsubsection{Ordering of Steady-state Values in the Homophily-Free Model}
In the homophily-free case with bias ($\lambda=0, b\ne 1/2$), the steady-state values $x^{*}_{j}$ satisfy
\begin{eqnarray*}
\frac{1+r_{j}}{1 + B Q_0(x^{*}_{j-1})} & = & x^{*}_{j}+\frac{r_{j}}{1 + B Q_{0}(x^{*}_{j})}\\
 & = & \frac{1}{1 + Q_0(x^{*}_{j})}+\frac{r_{j}}{1 + B Q_{0}(x^{*}_{j})},
\end{eqnarray*}
since $u=\frac{1}{1+Q_{0}(u)}$. Then if $b>\frac{1}{2}$ (bias favoring women), $B<1$, and 
\begin{eqnarray*}
\frac{1+r_{j}}{1 + B Q_0(x^{*}_{j-1})} & < & \frac{1}{1 + B Q_0(x^{*}_{j})}+\frac{r_{j}}{1 + B Q_{0}(x^{*}_{j})}\\
 & = & \frac{1+r_{j}}{1 + B Q_0(x^{*}_{j})},
\end{eqnarray*}
which means that $x^{*}_{j-1}<x^{*}_{j}$ for all $j$. Since $x^{*}_{0}=\frac{1}{2}$, this implies
\[
\frac{1}{2}<x^{*}_{1}<\ldots<x^{*}_{j}<\ldots<x^{*}_{L}.
\]
Similarly, $b<\frac{1}{2}$ (bias against women) implies
\[
x^{*}_{L}<\ldots<x^{*}_{j}<\ldots<x^{*}_{1}<\frac{1}{2}.
\]

\subsubsection{Degeneracy in the Bias-free Model}
\noindent\paragraph{Fullest Degeneracy for $r_{j} = 0$}
The bias-free model ($b=1/2$) has a symmetry related to our assumption of men and women behaving equally. Within a given hierarchy, the fractionation of men and women can be swapped, but the dynamics remain the same. This symmetry gives rise to the degenerate pitchfork bifurcations seen in Figure 7. The pitchfork bifurcation of fixed points becomes even more degenerate when $r_{j}\rightarrow0$.

For $r_{j}=0$, the fixed points necessarily satisfy $x_{j}^{*}=f_{j}^{*}$.
Note that for any value of $x_{j-1}^{*}$, $f_{j}^{*}$ is a shifted
sigmoid with a maximum slope of $\frac{\lambda}{4}$, and
\[
x_{j}^{*}=f_{j}^{*}=\frac{1}{1+\exp{\left[-\lambda\left(x_{j}^{*}-\left(x_{j-1}^{*}+\frac{\log Q_{0}\left(x_{j-1}^{*}\right)}{\lambda}\right)\right)\right]}}.
\]
For a given $x_{j-1}^{*}\in\left(0,1\right)$, there are at most three
solutions of $x_{j}^{*}=f_{j}^{*}$ for $\lambda>4$, and there is
exactly one solution for $\lambda\le4$. Since $L$ copies of this
equation are satisfied simultaneously, there are at most $3^{L}$
fixed points for $\lambda>4$ and exactly one for $\lambda\le4$.
Given $x_{0}^{*}=\frac{1}{2}$, the equi-gender state, $x_{j}^{*}=\frac{1}{2}\,\forall j$,
is a fixed point for all $\lambda$. Iso-gender states, $x_{j}^{*}=x_{\mathrm{iso}}^{*}$ for all $j>0$,
account for two more for $\lambda>4$. The equations for $j>1$ are trivially satisfied, and the fixed points are given implicitly by
the $j=1$ equation,
\[
x_{\mathrm{iso}}^{*}=\frac{1}{1+\mathrm{e}^{-\lambda\left(x_{\mathrm{iso}}^{*}-\frac{1}{2}\right)}}\implies\lambda=\frac{\log\frac{1-x_{\mathrm{iso}}^{*}}{x_{\mathrm{iso}}^{*}}}{\frac{1}{2}-x_{\mathrm{iso}}^{*}}\equiv\lambda_{\mathrm{iso}}\left(x_{\mathrm{iso}}^{*}\right),
\]
where $\lambda_{\mathrm{iso}}$ has even symmetry about $x_{\mathrm{iso}}^{*}=\frac{1}{2}$
such that there are two solutions, $x_{\mathrm{iso},+}^{*}>\frac{1}{2}$ and
$x_{\mathrm{iso},-}^{*}<\frac{1}{2}$. Finally, it can be shown that for $\lambda=\lambda_{\mathrm{iso}}\left(x_{\mathrm{iso},+}^{*}\right)=\lambda_{\mathrm{iso}}\left(x_{\mathrm{iso},-}^{*}\right)$,
there are additional fixed points where for each $j>0$, $x_{j}^{*}\in\{x_{\mathrm{iso},-}^{*},\frac{1}{2},x_{\mathrm{iso},+}^{*}\}$.
Since $\displaystyle \lim_{x_{\mathrm{iso}}^{*}\rightarrow\frac{1}{2}}\lambda_{\mathrm{iso}}\left(x_{\mathrm{iso}}^{*}\right)=4$,
all of these $3^{L}$ fixed points bifurcate from the equi-gender
fixed point at $\lambda=4$. The pitchfork
bifurcation is indeed highly degenerate in this case.

\noindent\paragraph{Stablity of the Equi- and Iso-gender Fixed Points}
The equi-gender and iso-gender fixed points as defined above persist for $r_{j}\ge0$, since for these states $f_{j}=x_{j}=f_{j+1}$ for all $j>0$. This further implies that $x_{j}=f_{j}=x^{*}$ for all $j>0$, where $x^{*}=\frac{1}{2}$ for $\lambda\le4$ or $x^{*}\in\{x_{\mathrm{iso},-}^{*},\frac{1}{2},x_{\mathrm{iso},+}^{*}\}$ for $\lambda=\lambda_{\mathrm{iso}}\left(x_{\mathrm{iso},+}^{*}\right)=\lambda_{\mathrm{iso}}\left(x_{\mathrm{iso},-}^{*}\right)>4$. The tridiagonal Jacobian matrix $M$ simplifies in this case. In row $j$, the lower, main, and upper diagonal entries are
\begin{align*}
m_{j,j-1} & = R_{j}(1+r_{j})\left(1-\mu\right)\!,\\
m_{j,j} & = R_{j} \left[(1+r_{j})\mu-1-r_{j}\left(1-\mu\right)\right]=R_{j} \left[\left(\mu-1\right)-r_{j}\left(1-2\mu\right)\right]\!,\\
m_{j,j+1} & = -R_{j}r_{j}\mu,
\end{align*}
where $\mu=\lambda x^{*}(1-x^{*})$. 

As an aside, we note a useful property of some tridiagonal matrices that can be used to bound their eigenvalues. Suppose that each pair of upper and lower diagonal entries in a tridiagonal matrix $T$ have opposite sign, i.e. $t_{i,j}t_{j,i}<0$ if $i \ne j$. Then one can choose a diagonal matrix $D$ such that the upper diagonal entries of $DTD^{-1}$ are the exact opposite of the lower diagonal entries. A similarity transformation by a diagonal matrix preserves the diagonal entries, and so the symmetric part of $DTD^{-1}$ is a diagonal matrix whose entries are the diagonal elements $t_{i,i}$ of $T$. Further, the eigenvalues of the symmetric part of a matrix bound the real parts of that matrix's eigenvalues. Consequently, the real parts of the eigenvalues of such a matrix $T$ are bounded
by the minimum and maximum diagonal entry.

For the Jacobian matrix $M$, we can use this bound if $0<\mu<1$. In that case, the fixed point is stable if $(\mu-1)-r_{j}(1-2\mu)<0$ for all $j$. Assuming $\mu\in\left(0,\frac{1}{2}\right)$, this requires that $\frac{\mu-1}{1-2\mu}<\min_{j}\{r_{j}\}$, which is always the case since $r_{j}\ge0$. Assuming instead that $\mu\in\left(\frac{1}{2},1\right)$, the fixed point is stable if $\frac{1-\mu}{2\mu-1}>\max_{j}\{r_{j}\}$, i.e. if $\mu<\frac{\max_{j}\{r_{j}\}+1}{2\max_{j}\{r_{j}\}+1}$. In the $r_{j}\rightarrow0$ limit, this, too, is a non-condition: the fixed point is stable for all $\mu\in(0,1)$. For $\max_{j}r_{j}\rightarrow\infty$, the condition simply requires $\mu<\frac{1}{2}$.

This has implications for the equi- and iso-gender fixed points. For the equi-gender fixed point, $\mu=\frac{\lambda}{4}$, and the fixed point is stable at least from $\lambda=0$ to $\lambda=2$, with the window of guaranteed stability increasing to $\lambda=4$ as $\max_{j}r_{j}\rightarrow0$.  The Gershgorin Circle Theorem guarantees that, in addition, it is stable for $\lambda<0$ (i.e., when people prefer to be among the opposite gender).
For the iso-gender fixed point, $\mu=\lambda_{\mathrm{iso}}\left(x^{*}_{\mathrm{iso}}\right)x^{*}_{\mathrm{iso}}\left(1-x^{*}_{\mathrm{iso}}\right)$, which lies in the interval $(0,1)$ for all $x^{*}_{\mathrm{iso}}\in(0,1)\setminus\left\{\frac{1}{2}\right\}$ and is $1$ for $x^{*}_{\mathrm{iso}}=\frac{1}{2}$. The fixed point is stable at least for $\mu<\frac{1}{2}$ or $\lambda$ greater than approximately $5.46798$. This is a transcendental number that does not depend on any parameter values, and it can be found by solving for $x^{*}_{\mathrm{iso}}$ numerically with $\mu=\frac{1}{2}$ and then computing $\lambda(x^{*}_{\mathrm{iso}})$. Its window of guaranteed stability increases with the lower limit approaching $\lambda=\lambda_{\mathrm{iso}}\left(\frac{1}{2}\right)=4$ as $\max_{j}\{r_{j}\}\rightarrow0$.

With these stability bounds in mind, we consider the oscillatory states to which the fixed points lose stability. We speculate that as $r_{j}\rightarrow0$, i.e.~the limit of highly bottom-heavy hierarchies, oscillatory solutions exist in smaller and smaller ranges of $\lambda$ around $\lambda=4$. On the other hand, if a hierarchy has more and more people in upper levels relative to lower ones, the oscillatory solutions do not necessarily dominate over an ever-expanding interval in $\lambda$; they likely appear at most for $\lambda$ from about $2$ to about $5.5$.

\subsubsection{Numerical Continuation for the Model with Bias and Homophily}
The above analysis of the fully degenerate pitchfork serves as starting point for numerical continuation of solution branches and bifurcation curves for the most general version of the model. Since we wish to continue from the degenerate $r_{j}=0$ situation, we introduce a new parameter $s$, such that $s=0$ gives the fully degenerate case and $s=1$ gives the desired $r_{j}$ values:
\begin{equation*}
\frac{dx_{j}}{dt}=R_{j}\left[(1+s r_{j})f_{j}-x_{j}-s r_{j}f_{j+1}\right].
\end{equation*}
Here, we detail our approach for the case when $L=3$, with parameter values as in Figure 7. We expect that the procedure generalizes well for different parameters. Starting with the $3^{3}=27$ fixed points at $s=0$ and $\lambda = \lambda_{\mathrm{iso}}(0.6)\approx 4.05$, we continue for increasing $s$. Only $11$ of these fixed points persist to $s=1$, those that appear for $\lambda\approx4.05$ in Figure 7, with the others being eliminated in saddle-node or pitchfork bifurcations. From the 11 fixed points, continuation of solution branches proceeds as is typical with AUTO. In particular, in generating Figure 7, we continue the fixed point branches in $\lambda$, noting any Hopf points and continuing limit cycle branches from there. For Figure 8, we continue the 11 fixed points first in $b$ down to $0.499$ and then in $\lambda$ as before. In this case, we also independently continue the upper stable limit cycle in $b$ (green band, Figures 7 and 8), since it is not tied to a fixed point branch via a Hopf bifurcation for $b=0.499$.

We find that for increasing $\lambda$, each limit cycle branch terminates in a homoclinic orbit. Additionally, it appears that as the period diverges there are a series of limit cycle saddle-node bifurcations occurring in a very small range of $\lambda$. While this phenomenon seems similar to that of a Shilnikov bifurcation, the saddle point associated with the homoclinic orbit is not a saddle-focus; it has purely real eigenvalues. While interesting dynamically, it occurs over a very small parameter range, with net result being a large-period termination of the limit cycle. We do not investigate further here.

\begin{figure}[p!]
	\begin{center}
		\includegraphics[trim=0.1in 0.03in 0.11in 0.03in,clip,width=0.88\textwidth]{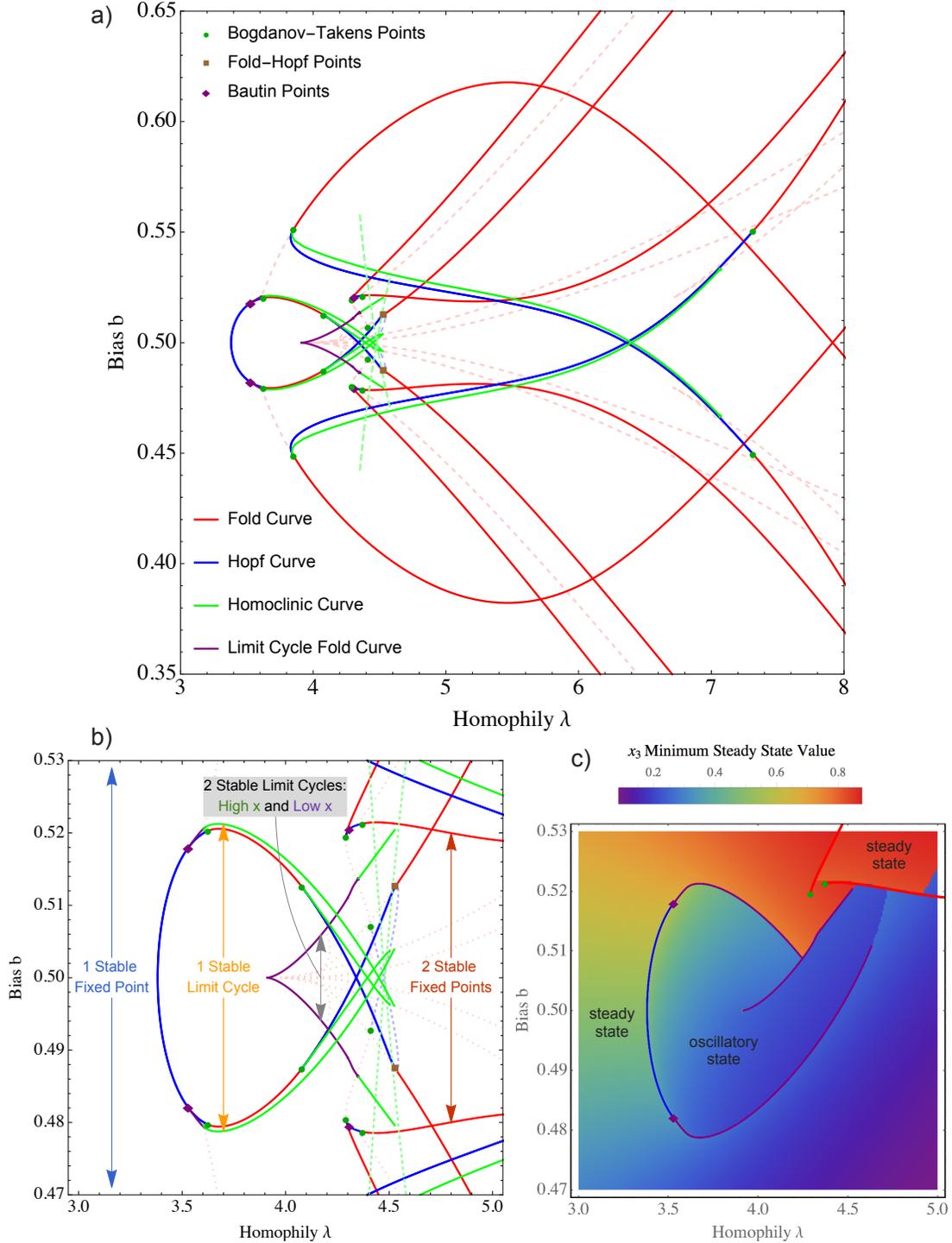}
	\end{center}
	\caption{Phase diagram across bias and homophily. a) Co-dimension 1 curves and co-dimension 2 points. The dashing indicates that the bifurcation involves no stable steady states. b) Zoom in on phase diagram, showing progression of steady states across homophily that persist for small bias. Arrows indicate the extents across bias of the stable states seen in Figures 7 and 8.  c) Minimum value that $x_{3}$ takes on in steady state for the male-dominated initial condition, $(x_{1}(0), x_{2}(0), x_{3}(0))=(0,0,0)$. The relevant bifurcation curves are overlaid and separate qualitatively different long-term behavior.}
	\label{fig:phaseDia}
\end{figure}

Continuation of the bifurcation points in the two parameter $\lambda-b$ space is more involved and proceeds in 4 stages:
\begin{enumerate}
\item From the 11 bias-free fixed points at $\lambda=4.5$ and the equi-gender fixed point at $\lambda=3.5$, we continue for $b\rightarrow 0$ and $b\rightarrow 1$, noting all co-dimension 1 points, here fold points and Hopf bifurcations.
\item We continue each type of bifurcation in $\lambda$ and $b$, making note of all co-dimension 2 points. Here we find Bogdanov-Takens, fold-Hopf, and Bautin points.
\item In the cases of the Bogdanov-Takens and Bautin points, limit-cycle bifurcation curves emanate from the co-dimension 2 point generically. They are homoclinic curves and limit cycle fold curves. We approach these via the associated Hopf bifurcations, continuing the limit cycle in each case along an ad-hoc path in $\lambda$ and $b$. We then continue the limit cycle fold and homoclinic curves. For the homoclinic curves, we use the HomCont functionality built into AUTO by obtaining an initial approximate homoclinic solution from a large period limit cycle with a period of at least $\mathcal{O}(10^{6})$.
\item Lastly, we capture the unfolding of the pitchfork bifurcation of limit cycles, which occurs at a cusp point of limit cycle fold curves. By again ad-hoc paths in $\lambda$ and $b$, we approach the fold curves for $b\ne 1/2$ and then continue. 
\end{enumerate}
The results of this procedure are summarized in a phase diagram (Figure \ref{fig:phaseDia}). Due to numerical difficulties, some of the homoclinic curves are cut short. Also, note that at this scale some of the limit cycle saddle-node and homoclinic curves seem to coincide, reflecting the Shilnikov-like dynamics. Using this phase diagram, we extend our understanding of model dynamics from the bias-free case to the case with some bias (Figure \ref{fig:phaseDia}b). For small biases, $b$ between approximately $0.48$ and $0.52$, the qualitative progression of stable states is similar to that in the bias-free model: single ``equi-gender-like'' fixed point $\rightarrow$ oscillations $\rightarrow$ two male- or female-dominated ``iso-gender-like'' fixed points (cf.~Figures 7 and 8). Note

\noindent that we do not in general rule out the possibility of other stable states that may appear yet higher homophily values. Such homophily values, however, are likely to be physically unrealistic.

We also investigate which of the many states reflected in the phase diagram are physically relevant. For a historically male-dominated hierarchy, the dynamics evolving from an initial condition of $x_{j}=0$ are perhaps the most relevant. Across bias and homophily, the long-term states show moderate deviations from gender parity, oscillatory states, and a highly polarized state (Figure \ref{fig:phaseDia}c). For the most part, the regions showing qualitatively different behavior are separated by bifurcation curves. In those cases, the results will not change for sufficiently small perturbations of the initial condition.
In a few instances, though, a qualitative change in the observed state occurs when the separatrix between two stable states passes over the initial state, $(0,0,0)$.

\subsubsection{Oscillations and Single-gender-dominated States\label{subsec:Oscillations-and-IsoGender}}

The emergence of oscillations and of single-gender-dominated states
for low or no bias may not be intuitively obvious. In brief, those
states occur for moderate and high levels of homophily, respectively,
and are due to a balance between homophilic pressures and the availability
of promotable men/women. Here we detail the effect those influences
have on the 2-layer dynamics depicted in Figure \ref{fig:equi+oscillations+iso}.

For each low, moderate, and high homophily, Figure \ref{fig:equi+oscillations+iso}
shows snapshots in time of $x_{1}$, $x_{2}$, $f_{1}$ and $f_{2}$
along with the instantaneous rates of change in $x_{1}$ and $x_{2}$.
$f_{1}$ is a one-to-one function of $x_{1}$, as accounted for by
the gray lines in each snapshot, and $x_{2}$ (generally $x_{L}$)
simply follows $f_{2}$, as indicated by the blue dashed line. The
dynamics of $x_{2}$ and $f_{1}$ are therefore secondary to that
of $x_{1}$ and $f_{2}$, and we can foucs primarily on $x_{1}$ and
$f_{2}$ and appeal to the gray and blue lines for a complete dynamical
understanding. Recall that the gender fractionation $f_{2}$ of the
flow of men and women promoted from level $1$ to level $2$ is given
by $f_{2}=\left(1+Q_{0}\left(x_{1}\right)Q\left(x_{2}-x_{1}\right)\right)^{-1}$.
It depends both on the number of men/women available in level $1$
to be promoted and on the homophily those men/women feel. We find
it useful to consider those two factors separately in two auxiliary
forms of $f_{2}$:
\begin{enumerate}
\item considering only the availability of men/women in level $1$,
\[
f_{2}^{\text{availability}}=\frac{1}{1+Q_{0}\left(x_{1}\right)}=x_{1},
\]
i.e. what the flow fractionation would be without homophily. This
value is plotted using small red dots in Figure \ref{fig:equi+oscillations+iso}.
\item considering only the homophily felt by the men/women in level $1$,
\[
f_{2}^{\mathrm{homophily}}=\frac{1}{1+Q\left(x_{2}-x_{1}\right)},
\]
i.e. what the flow fractionation would be with identical availability
of men and women but with the actual homophilic pressure. This value
is plotted using small yellow dots in Figure \ref{fig:equi+oscillations+iso}.
\end{enumerate}
The actual flow $f_{2}$ will be influenced by both, but it is often
dominated by one. In those cases, the blue $f_{2}$ dot is ``pulled''
strongly towards either the small yellow $f_{2}^{\mathrm{homophily}}$
dot or the small red $f_{2}^{\text{availability}}$ dot. We now describe
how the influences of $f_{2}^{\mathrm{homophily}}$ and $f_{2}^{\text{availability}}$
result in each of the different types of final states: states near
gender parity, oscillations, and single-gender dominated states.

For low homophily (Figure \ref{fig:equi+oscillations+iso}a), gender
parity is achieved in the long run, and the dominant contribution
to $f_{2}$ is always $f_{2}^{\text{availability}}$. This means that
the gender fractionation will naturally equilibrate as it does in
the null model: $x_{2}$ tends towards $x_{1}$, and $x_{1}$ towards
$x_{0}=\frac{1}{2}$. A detailed description of each snapshot in Figure
\ref{fig:equi+oscillations+iso}a follows:
\begin{itemize}
\item[i)] $x_{1}$ is decreasing because more women are being promoted to
level $2$ than are coming in from the general population. $x_{2}$
is tending towards $x_{1}$, because the promotion flow into level
$2$ has roughly the same fractionation as level $1$.
\item[ii)] $x_{2}$ reaches a maximum as $x_{1}$ decreases past it.
\item[iii)] $x_{1}$ continues to decrease and converges to $\frac{1}{2}$.
$x_{2}$ follows.
\item[iv)] Gender parity is reached.
\end{itemize}
For moderate homophily (Figure \ref{fig:equi+oscillations+iso}b),
oscillations can form because $f_{2}^{\mathrm{homophily}}$ periodically
becomes the dominant contribution to $f_{2}$. Here, the homophily
is strong enough that a prevalence of woman at level $2$ can sometimes
disproportionally incentivize applications and draw women from level
$1$ (Figure \ref{fig:equi+oscillations+iso}b.iv). But it not sufficiently
strong to always replace the promoted women with women from the general
population. This reduces the availabilty of women that can be promoted
later, and the level $2$ will eventually see a decrease in women
(Figure \ref{fig:equi+oscillations+iso}b.v). This in turn will create
a homophilic suppression of female applicants, resulting in an increase
in women at the level $1$. Eventually, though, the lower level will
become so female-dominated that there are fewer male applicants available
to be promoted to the level $2$. This then causes an increase in
the fraction of women at level $2$ (Figure \ref{fig:equi+oscillations+iso}b.i)
and completes the cycle. 

Note that even though the snapshot Figure \ref{fig:equi+oscillations+iso}b.iii
is qualitatively similar to Figure \ref{fig:equi+oscillations+iso}a.ii
for low homophily, $x_{1}$ and $x_{2}$ do not converge to gender
parity in this case. Instead as $x_{1}$ approaches $\frac{1}{2}$
(Figure \ref{fig:equi+oscillations+iso}b.iv), $f_{2}^{\mathrm{homophily}}$
dominates over $f_{2}^{\text{availability}}$ reflecting the fact
that promotion to level $2$ is quite attractive to women in level
$1$. This causes an exodus of women from level $1$, pushing $x_{1}$
further down and across $\frac{1}{2}$. A detailed description of
each snapshot in Figure \ref{fig:equi+oscillations+iso}b follows:
\begin{itemize}
\item[i)] $x_{2}$ is initially increasing because of the good availability
of women to be promoted from level $1$. $x_{1}$ is also increasing
because level $1$ is attractive to women in the general population
due to homophily.
\item[ii)] $x_{1}$ reaches a maximum as the homophily-driven net flow of
women into level $1$ is overcome by the availability-driven flow
into level $2$.
\item[iii)] $x_{2}$ reaches a maximum as $x_{1}$ and the availability-driven
flow $f_{2}$ decreases past it.
\item[iv)] $f_{2}$ becomes homophily-dominated, pushing $x_{1}$ down past
$\frac{1}{2}$ as women are promoted to level $2$. 
\item[v)] $x_{1}$ has reached a minimum, and the system is at state roughly
symmetric to that in Figure \ref{fig:equi+oscillations+iso}b.ii with
men and women swapped. The oscillation will continue with $x_{1}$
increasing to a maximum and the cycle repeating. 
\end{itemize}
For high homophily (Figure \ref{fig:equi+oscillations+iso}c), a female-dominated
state emerges as the dynamics proceed in two phases: first $f_{2}^{\mathrm{homophily}}$
dominates and then $f_{2}^{\text{availability}}$ does. Homophily
is so strong here that, since level $1$ initially has female representation
a little greater than the general populaiton, it quickly becomes very
dominated by women (Figure \ref{fig:equi+oscillations+iso}c.ii).
Eventually, practically no men are available to be promoted from level
$1$ (Figure \ref{fig:equi+oscillations+iso}c.iii), and level $2$
subsequently follows in becoming female-dominated (Figure \ref{fig:equi+oscillations+iso}c.iv).
A detailed description of each snapshot follows:
\begin{itemize}
\item[i)] The strong homophily means the level $1$ is very attractive to
women in the general population, but level $2$ is unattractive to
those already in level $1$.
\item[ii)] $x_{1}$ increases very rapidly and $x_{2}$ decreases.
\item[iii)] As $x_{1}$ approaches $1$, the are very few men left to be
promoted to level $2$. The scarcity overwhelms the homophily-driven
flow of men to level $2$, and $x_{2}$ starts to increase.
\item[iv)] Availability remains the dominant force, pulling $x_{2}$ up towards
$x_{1}$.
\item[v)] The female-dominated iso-gender state is reached.
\end{itemize}

\begin{figure}[p!]
	\begin{centering}
		\includegraphics[trim=0.22in 0.074in 0.14in 0.1in,clip]{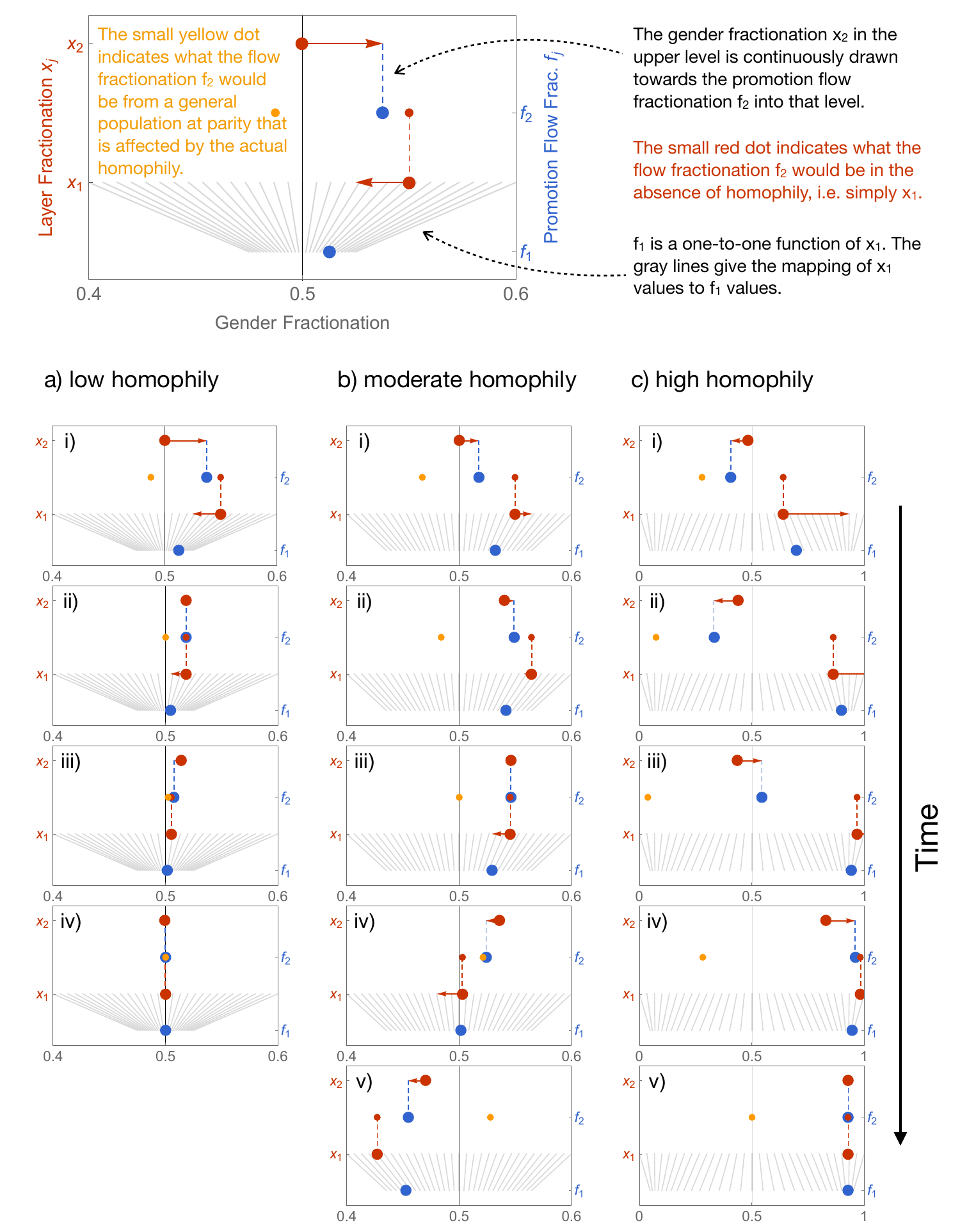}
	\end{centering}
	\caption{Qualitative depiction of snapshots in the time evolution for low,
moderate and high homophily and no bias, resulting in the long run
in gender parity, oscillations in gender fractionation, and a state
dominated by one gender (the iso-gender fixed point), respectively.
For illustration purposes, we choose $L=2$, $R_{2}=0.8$, $r_{1}\rightarrow\infty$
and $R_{1}\rightarrow0$ such that $R_{1}r_{1}\rightarrow1$ (giving
$\dot{x}_{1}=f_{1}-f_{2}$), and an initial condition of $\left(x_{1},x_{2}\right)=\left(0.55,0.5\right)$.
We expect that the qualitative behaviors depicted here reflect those
for other parameter choices. The red arrows indicate the direction
and magnitude (not to scale) of motion of $x_{1}$ and $x_{2}$, and
the text surrounding the topmost plot describes the meaning of the
other graphical components.\label{fig:equi+oscillations+iso}
	}
	\label{fig:equi+oscillations+iso}
\end{figure}

For high homophily (Figure \ref{fig:equi+oscillations+iso}c), the
dynamics proceed in two phases. First $f_{2}^{\mathrm{homophily}}$dominates,
making level $1$ very female-dominated and level $2$ moderately
male-dominated. Then $f_{2}^{\text{availability}}$ dominates as the
availability of men in level $1$ dwindles. As a result, the number
of women in level $2$ increases to meet the fractionation value in
level $1$: the iso-gender fixed point. A detailed description of
each snapshot follows:
\begin{itemize}
\item[i)] The strong homophily means the level $1$ is very attractive to
women in the general population, but level $2$ is unattractive to
those already in level $1$.
\item[ii)] $x_{1}$ increases very rapidly and $x_{2}$ decreases.
\item[iii)] As $x_{1}$ approaches $1$, the are very few men left to be
promoted to level $2$. The scarcity overwhelms the homophily-driven
flow of men to level $2$, and $x_{2}$ starts to increase.
\item[iv)] Availability remains the dominant force, pulling $x_{2}$ up towards
$x_{1}$.
\item[v)] The female-dominated iso-gender state is reached.
\end{itemize}

\subsection{Fitting Algorithm}
\begin{figure}[htb]
  \begin{center}
    \includegraphics[width=0.9\textwidth]{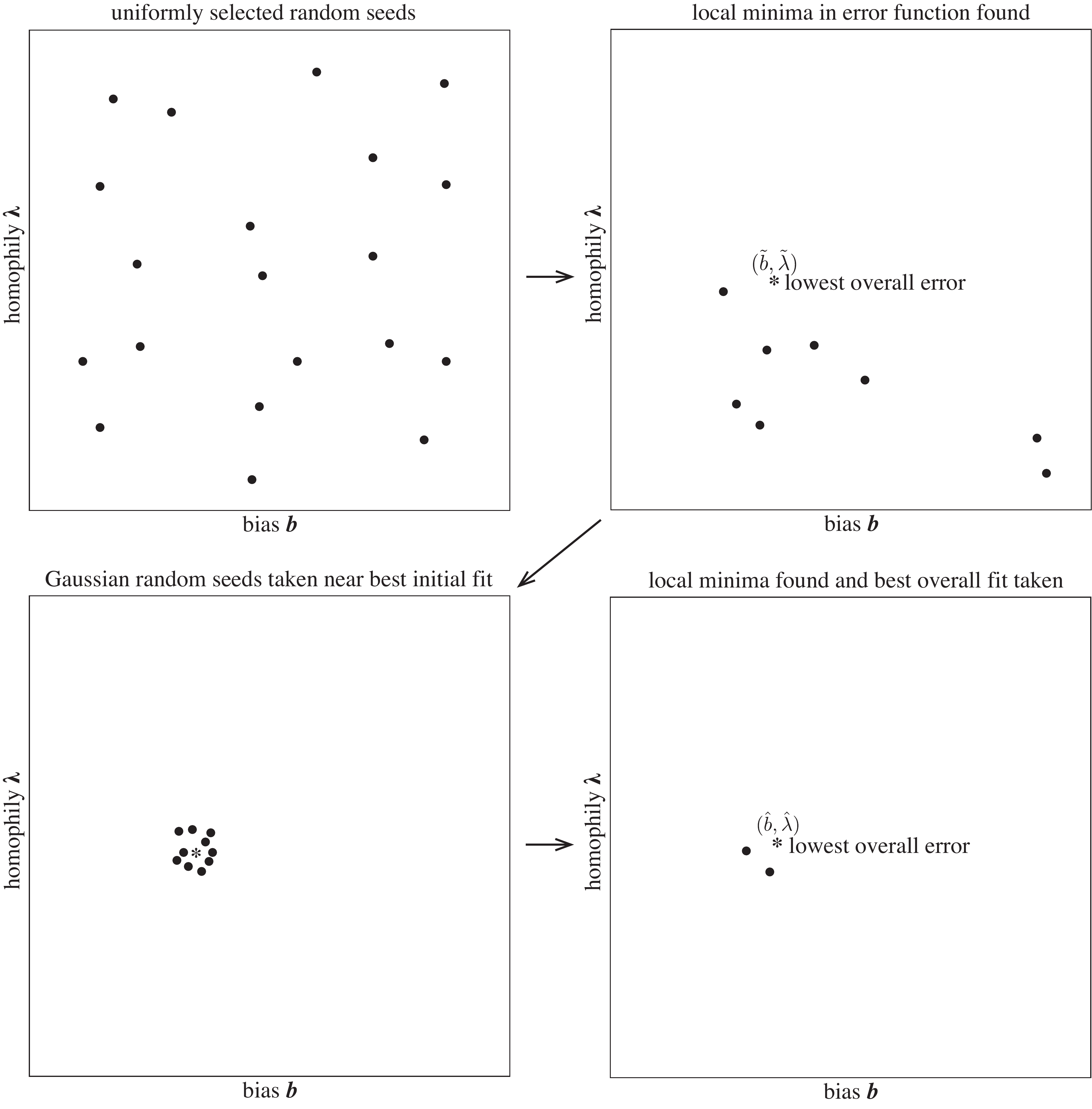}
  \end{center}
  \caption{Illustration of the model fitting algorithm. We seed the fitting algorithm with 20 initial guesses for the fitting parameters $b$ and $\lambda$, selected uniformly (top left). We then run the Nelder-Mead error minimization algorithm to find a set of local minima, one of which will have the absolute smallest error (top right). We then seed the fitting algorithm with 10 guess for the fitting parameters $b$ and $\lambda$, selected normally about the previous best fit (bottom left). Finally, we run the Nelder-Mead error minimization algorithm to find a set of local minima, one of which will have the smallest absolute error (bottom right). We take this to be our best fit.}
  \label{fig:fitdemo}
\end{figure}

See Figure \ref{fig:fitdemo} for a pictorial representation of the model fitting algorithm.

\subsection{Model Fitting Results}
\begin{figure}[htb]
  \begin{center}
    \includegraphics[width=0.8\textwidth]{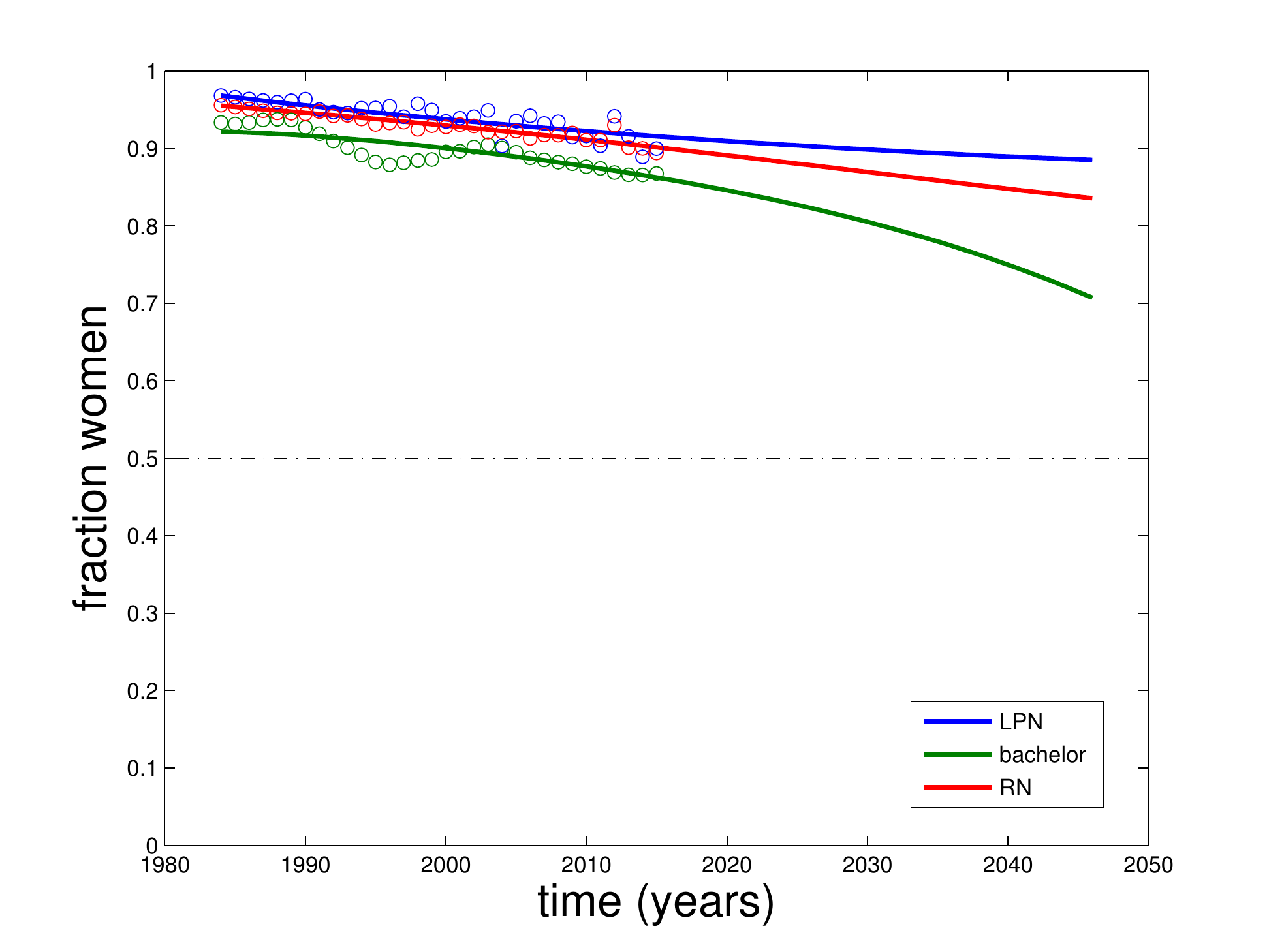} \\ \includegraphics[width=0.8\textwidth]{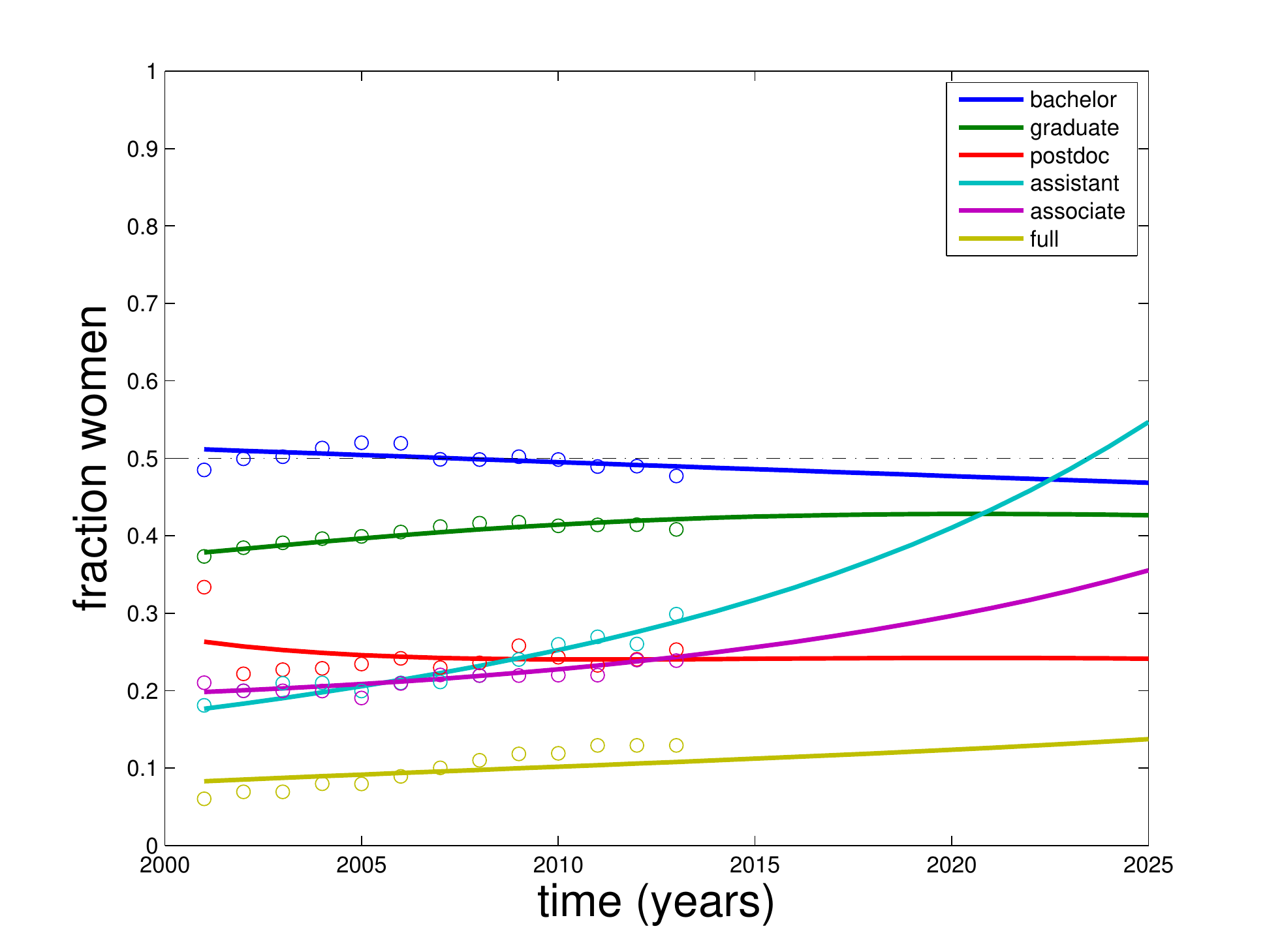}
  \end{center}
  \caption{Model fit to data from nursing\cite{ies2017nces,brotherton2017graduate,brotherton2016graduate,brotherton2015graduate,msu2017cihws,uscb2013nurse,hrsa2008nurse} (top) and academic chemistry\cite{nsf2017nss,ies2017nces,acs2015cen} (bottom).}
  \label{fig:bestfit1}
\end{figure}

\begin{figure}[htb]
  \begin{center}
    \includegraphics[width=0.8\textwidth]{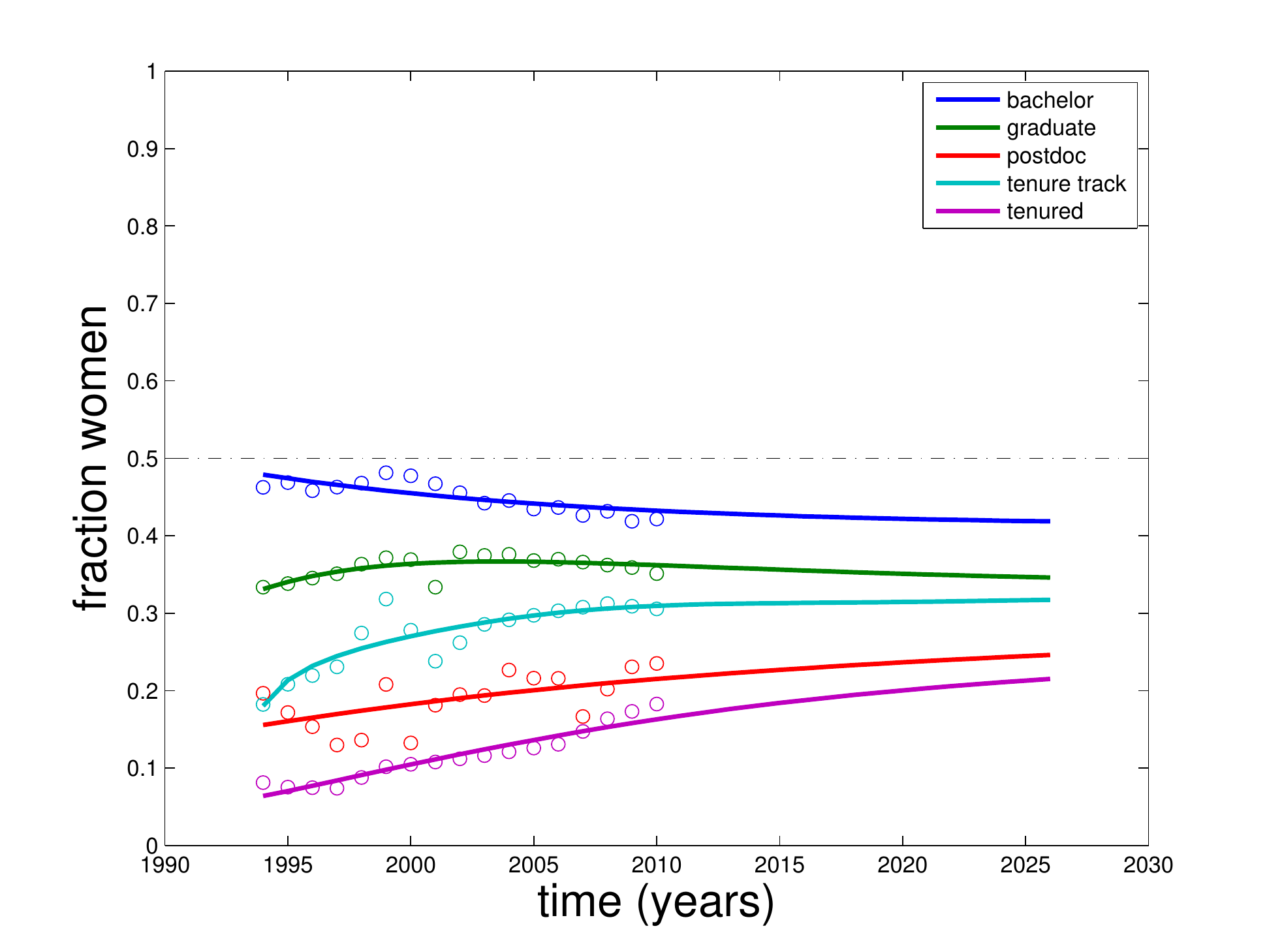} \\ \includegraphics[width=0.8\textwidth]{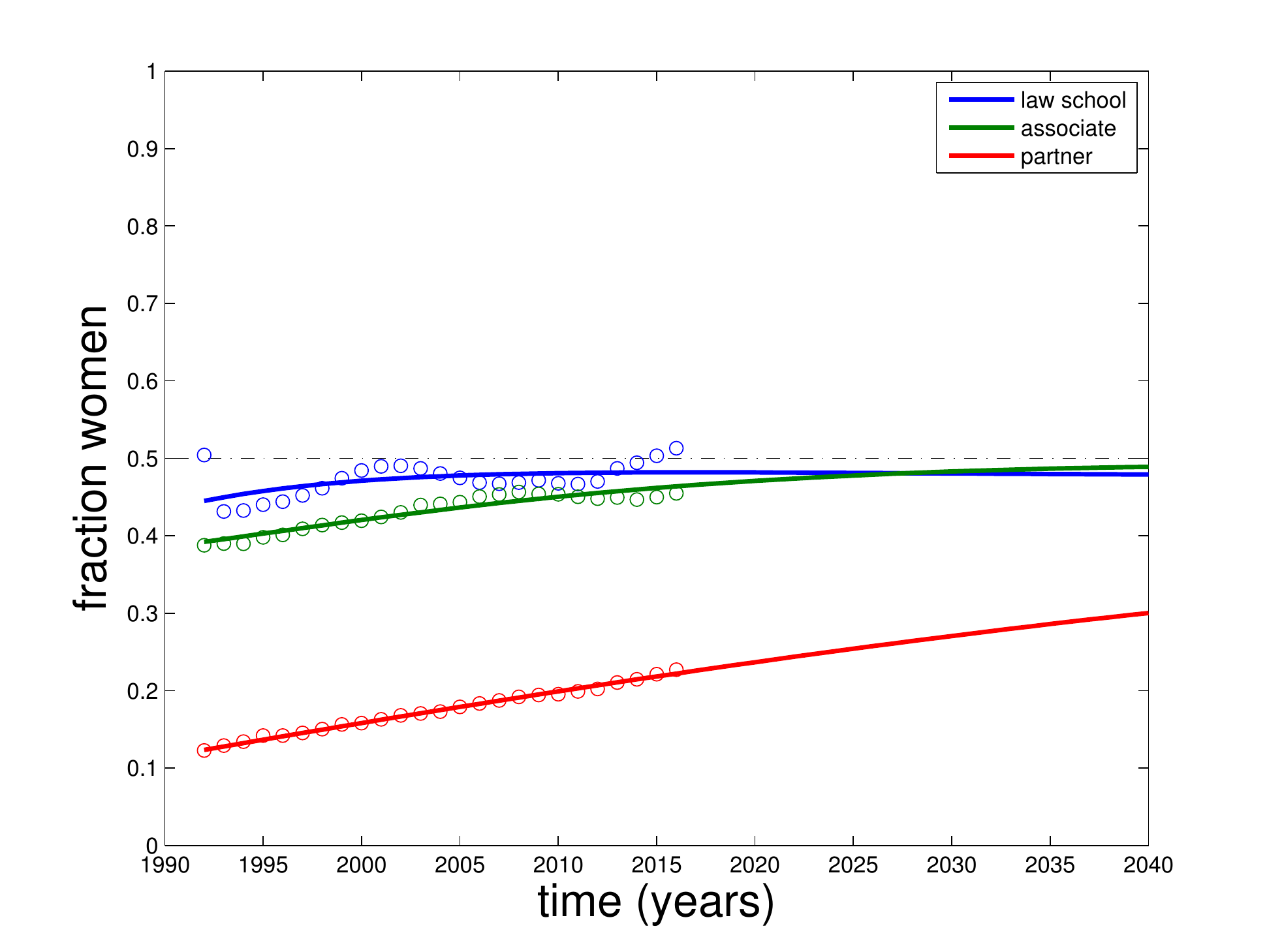}
  \end{center}
  \caption{Model fit to data from academic math/statistics\cite{nsf2017ncses,nsf2017nss,ies2017nces} (top) and law\cite{aba2013law,nalp2018law,nawl2017law} (bottom).}
  \label{fig:bestfit2}
\end{figure}

\begin{figure}[htb]
  \begin{center}
    \includegraphics[width=0.8\textwidth]{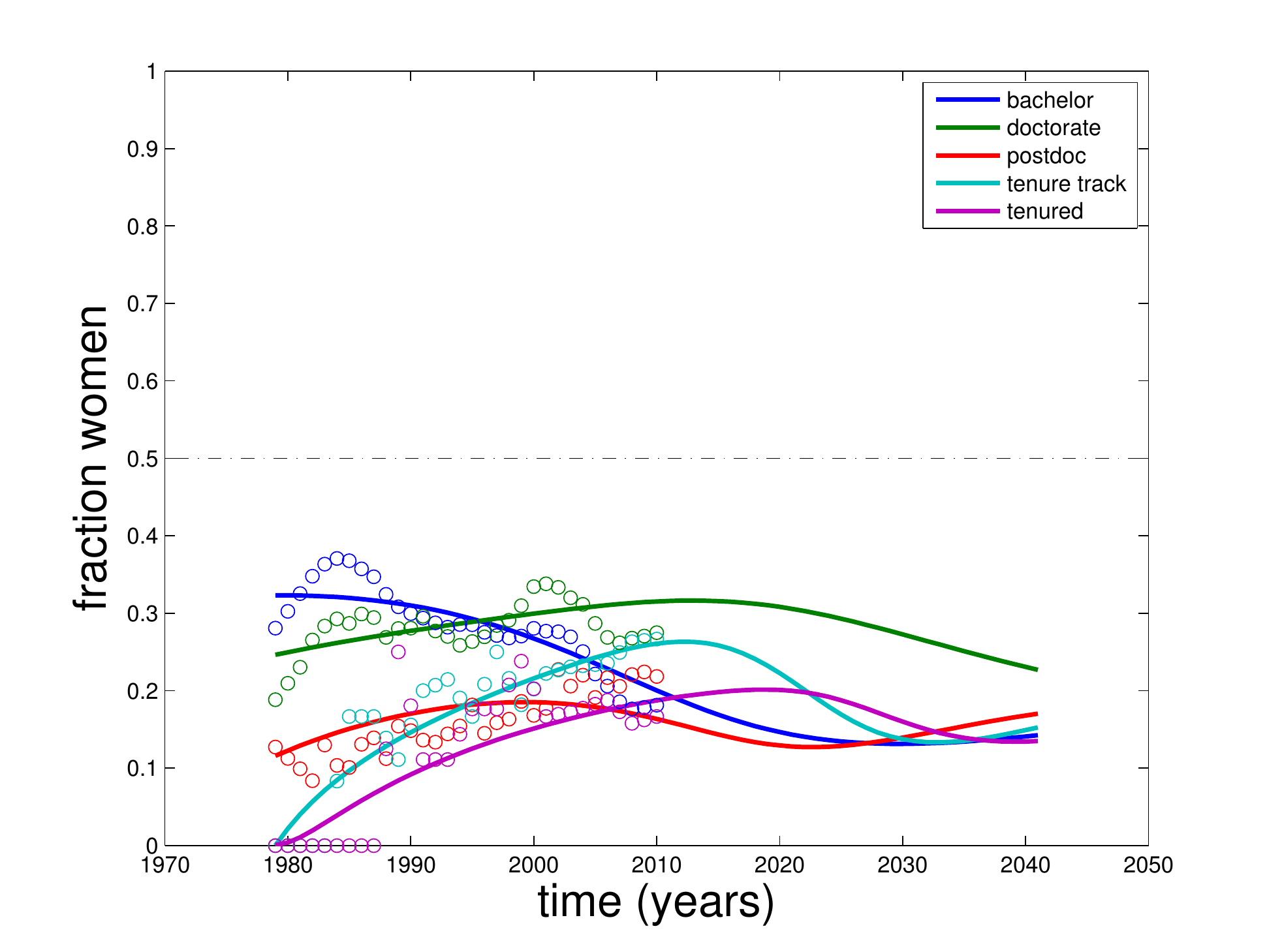} \\ \includegraphics[width=0.8\textwidth]{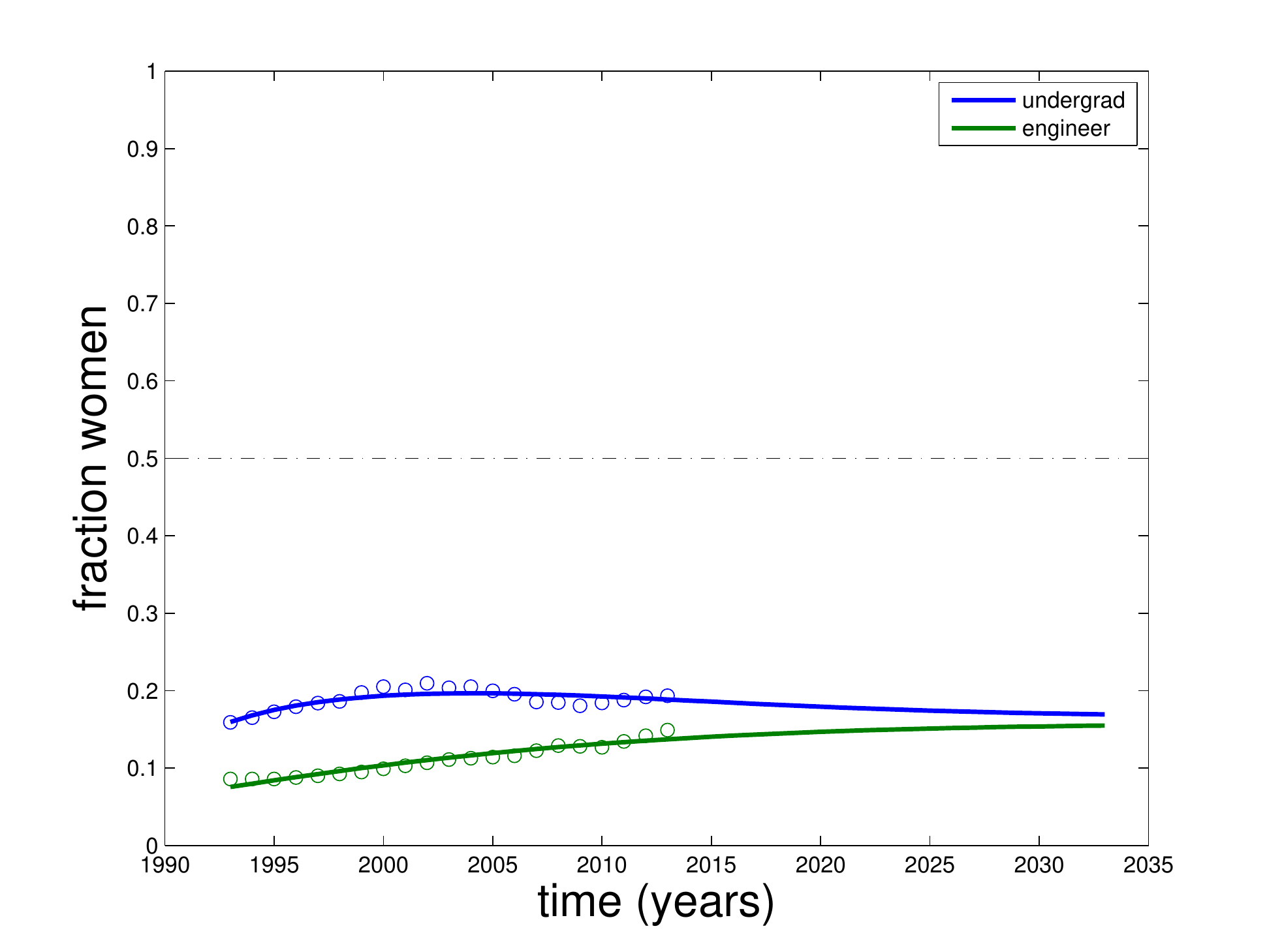}
  \end{center}
  \caption{Model fit to data from academic computer science\cite{nsf2017ncses,nsf2017nss,ies2017nces} (top) and engineering practice\cite{nsf2017ncses,ies2017nces} (bottom).}
  \label{fig:bestfit3}
\end{figure}

\begin{figure}[htb]
  \begin{center}
    \includegraphics[width=0.8\textwidth]{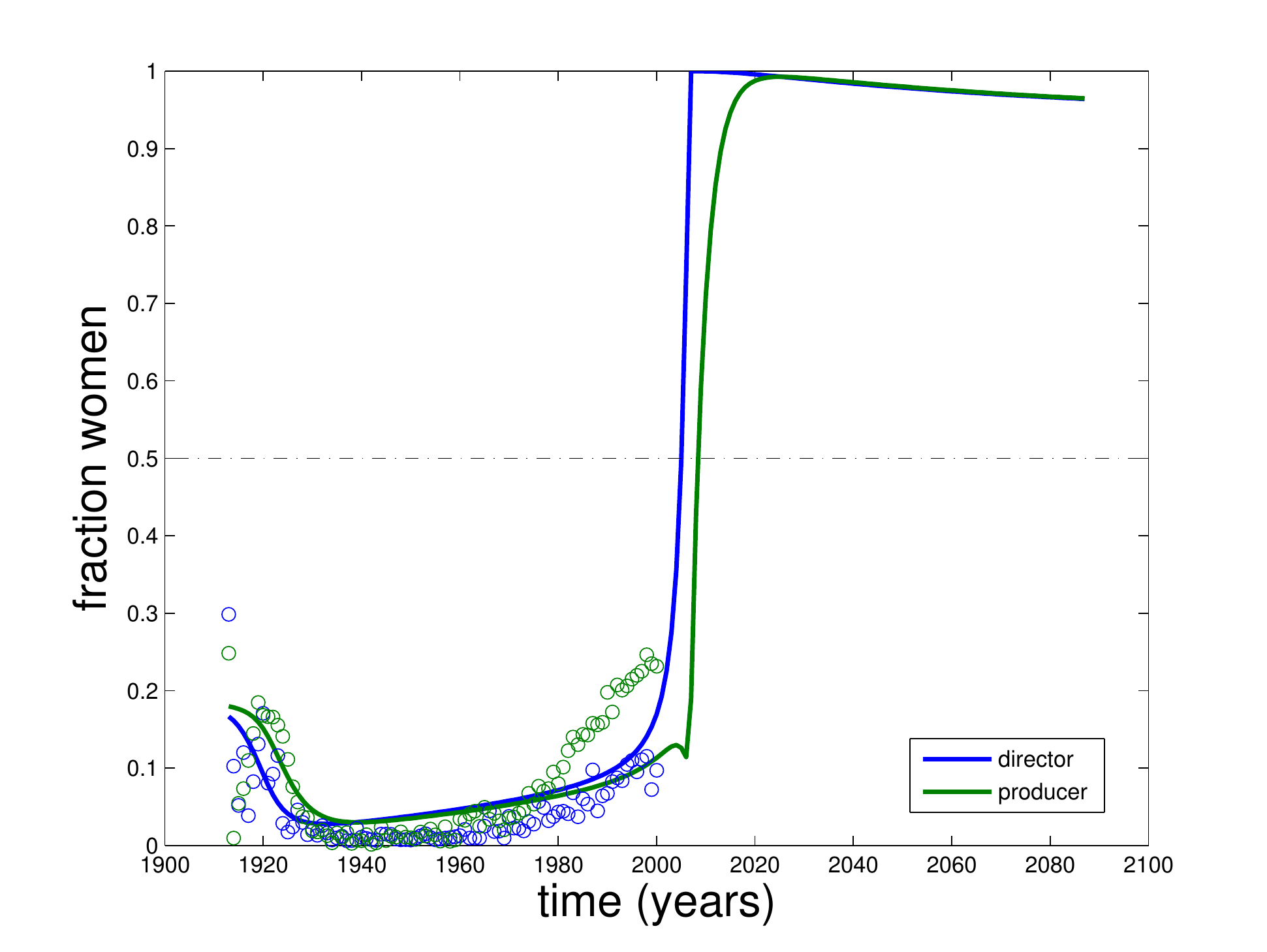} \\ \includegraphics[width=0.8\textwidth]{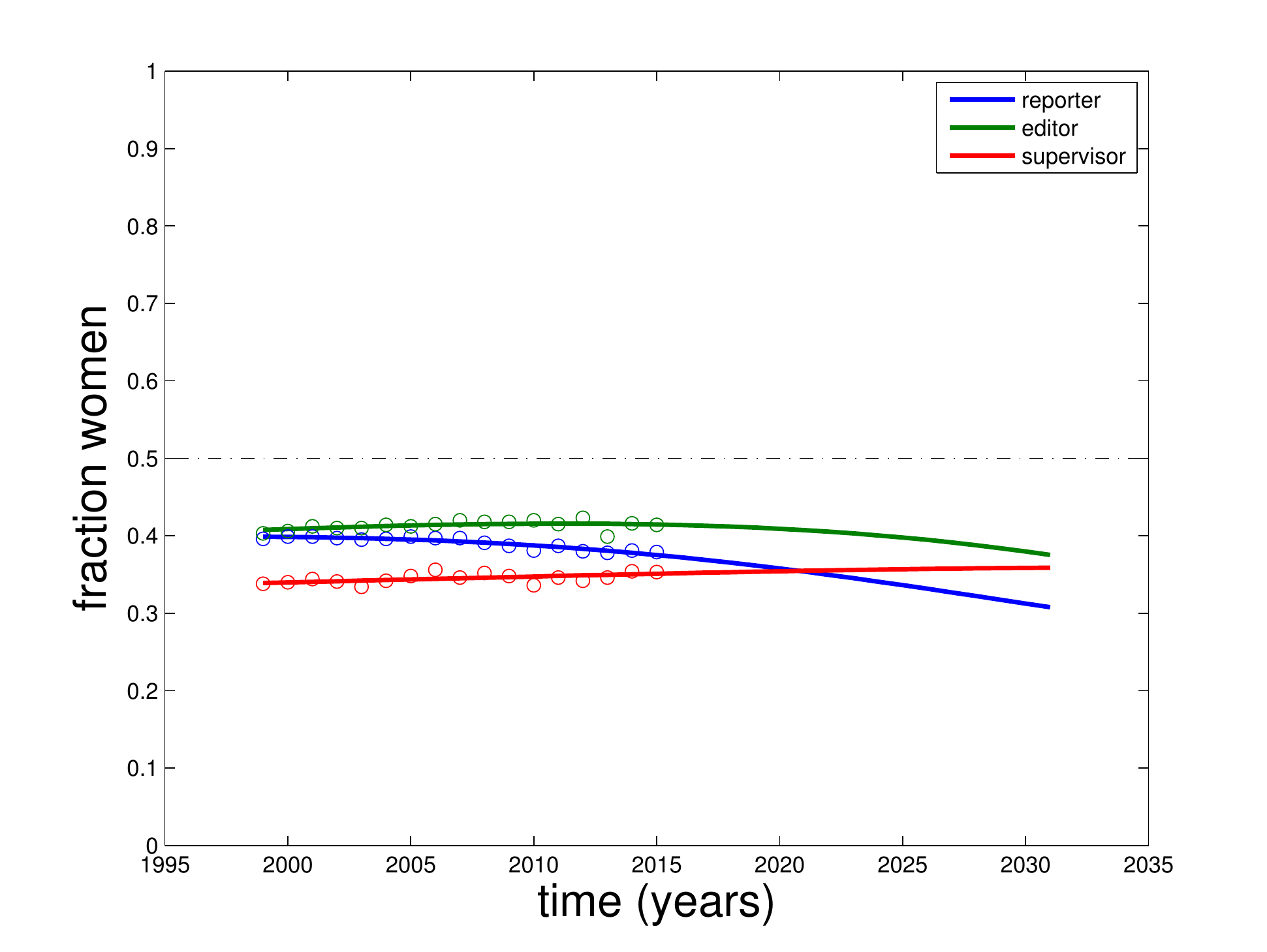}
  \end{center}
  \caption{Model fit to data from film\cite{lab_2017} (top) and journalism\cite{asne2015} (bottom).}
  \label{fig:bestfit4}
\end{figure}

\begin{figure}[htb]
  \begin{center}
    \includegraphics[width=0.8\textwidth]{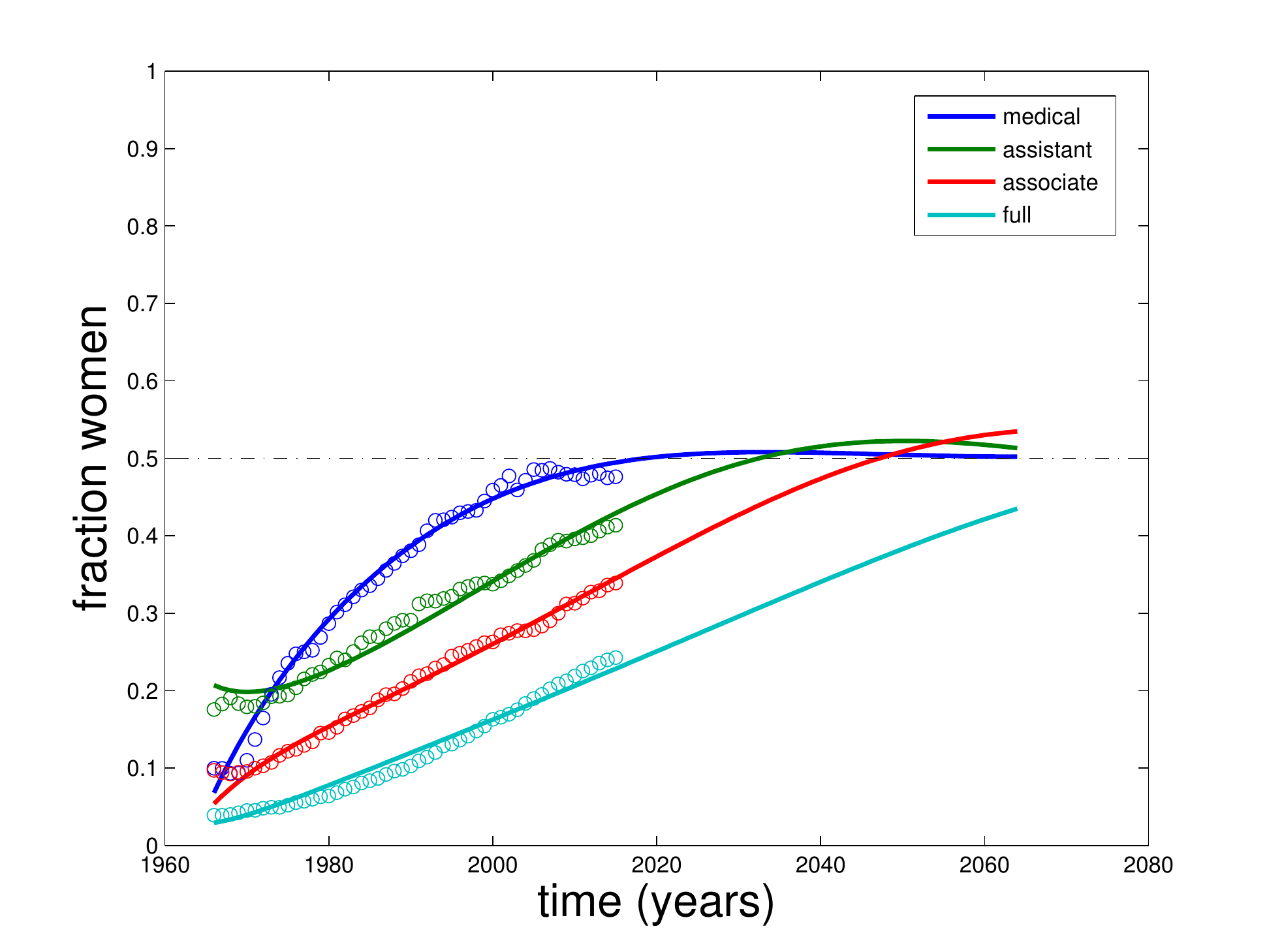} \\ \includegraphics[width=0.8\textwidth]{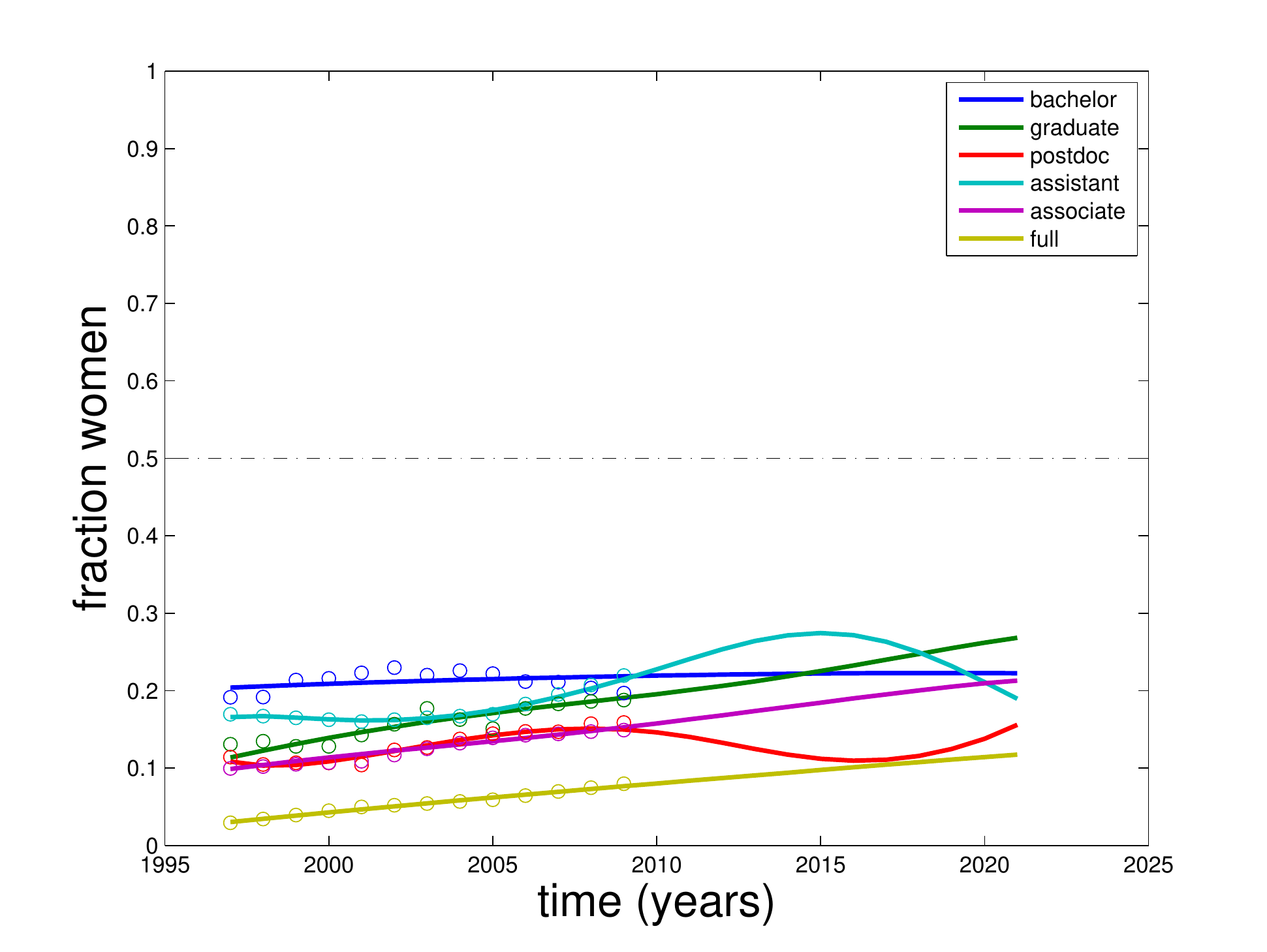}
  \end{center}
  \caption{Model fit to data from academic basic science medicine\cite{kff2017med,abim2018med,aamc2016med} (top) and academic physics\cite{nsf2017ncses,nsf2017nss,ies2017nces,nasem2013seeking} (bottom).}
  \label{fig:bestfit5}
\end{figure}

\begin{figure}[htb]
  \begin{center}
    \includegraphics[width=0.8\textwidth]{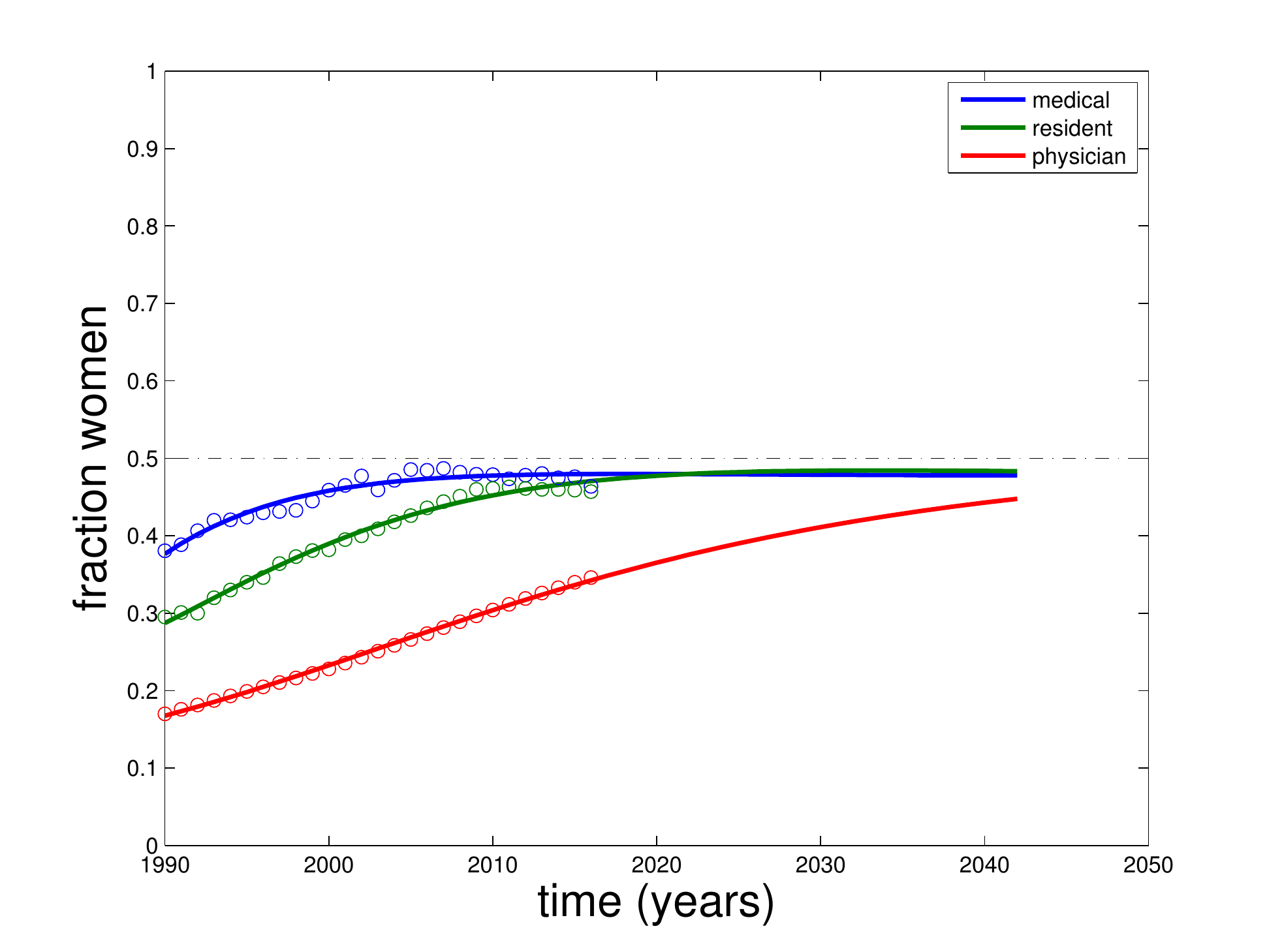} \\ \includegraphics[width=0.8\textwidth]{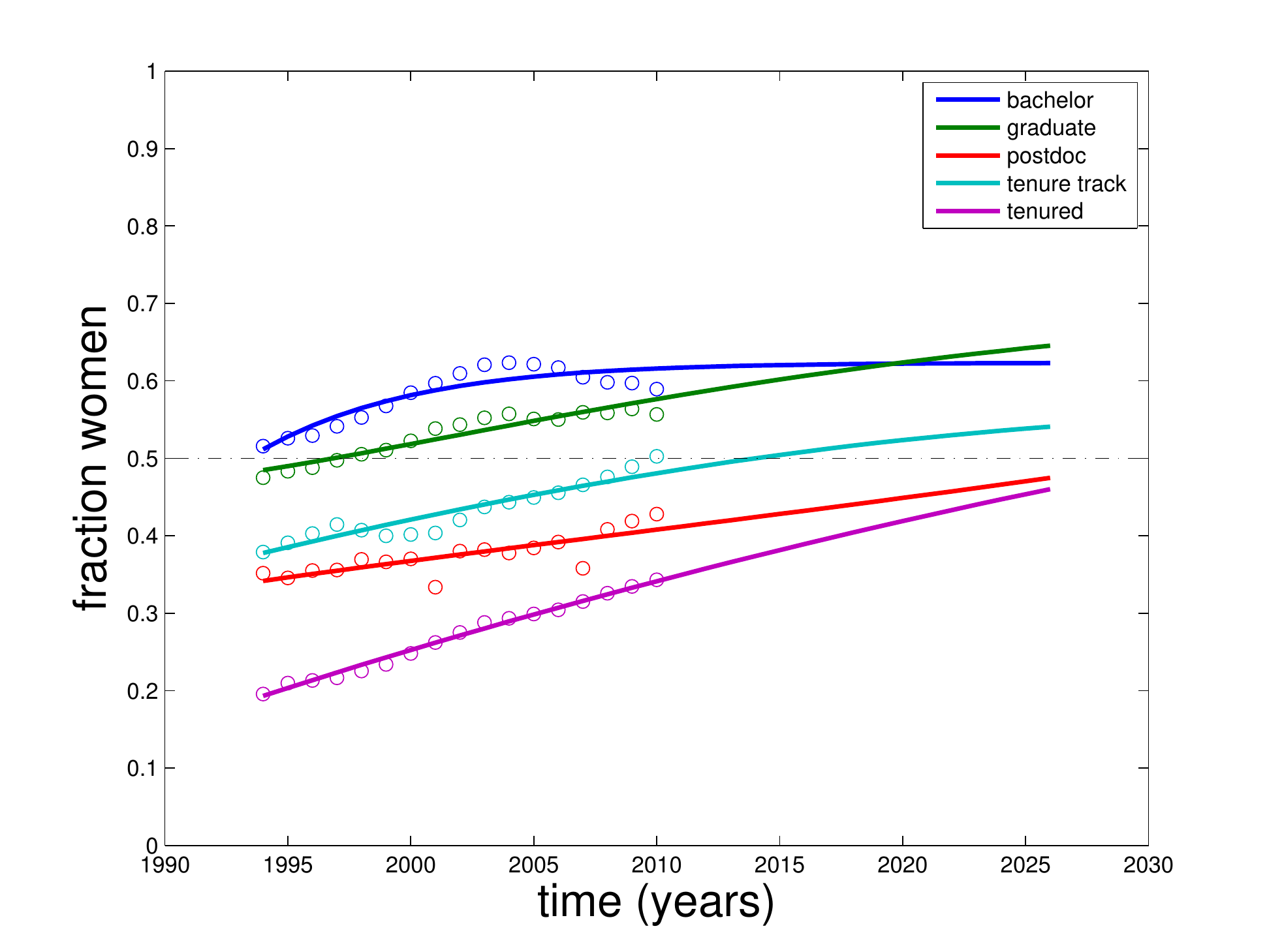}
  \end{center}
  \caption{Model fit to data from medical practice\cite{kff2017med,abim2018med,ies2017nces} (top) and academic biology\cite{nsf2017ncses,nsf2017nss,ies2017nces} (bottom).}
  \label{fig:bestfit6}
\end{figure}

\begin{figure}[htb]
  \begin{center}
    \includegraphics[width=0.8\textwidth]{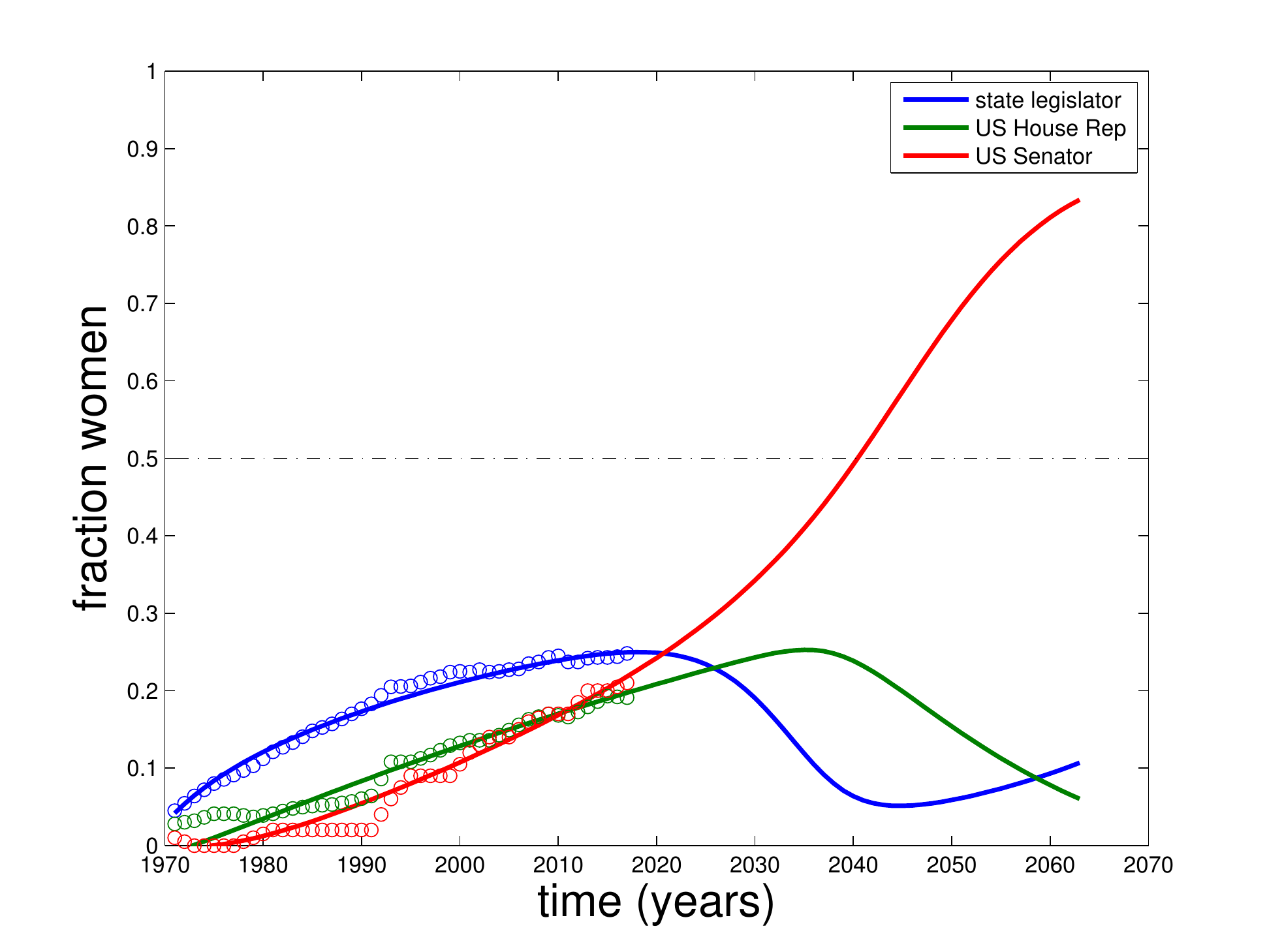} \\ \includegraphics[width=0.8\textwidth]{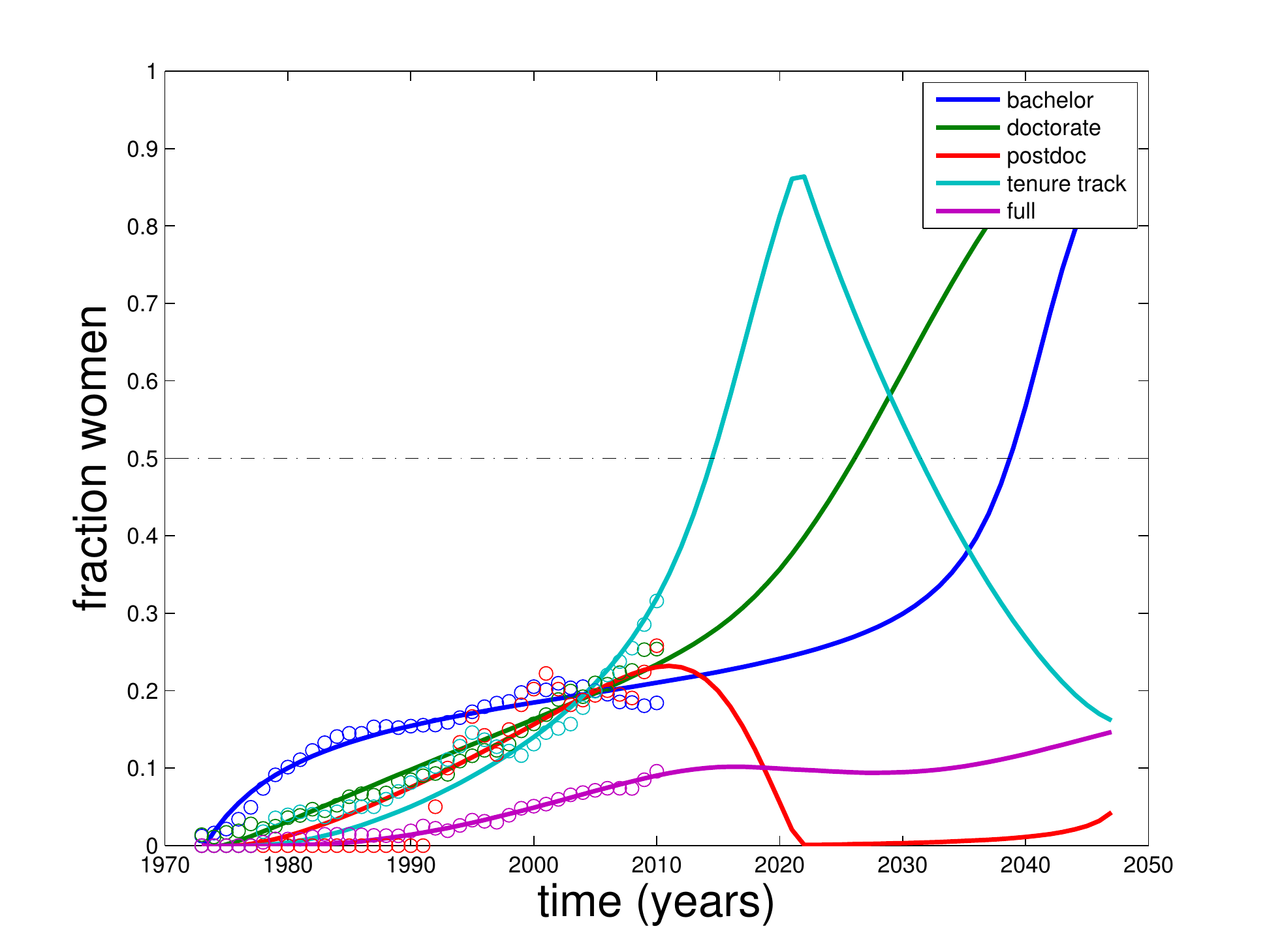}
  \end{center}
  \caption{Model fit to data from politics\cite{nawl2018rut,pew2018gov} (top) and academic engineering\cite{nsf2017ncses,nsf2017nss,ies2017nces} (bottom).}
  \label{fig:bestfit7}
\end{figure}

All plots of data with best fits not displayed in the main paper are shown in Figures \ref{fig:bestfit1}-\ref{fig:bestfit7}.

All best fit bias $b$ and homophily $\lambda$ parameters are reproduced in Table \ref{tab:results}. To compare with the model by Holman et al. \cite{holman2018gender}, we also compute the predicted number of years from 2018 until gender parity is reached for both the top and bottom levels. This is not an exact replication of Holman's study because we are not using last and first authorship as an indicator of top and bottom levels, respectively. Like Holman's model predictions, we see a wide range of times to parity; unlike their model, our model predicts some fields will never reach gender parity.

\begin{table}[htb]
\begin{center}
  \begin{tabular}{| p{5.5cm}  x{2.3cm}  x{2.6cm} x{2.7cm} @{\hskip 0\tabcolsep} x{2.7cm} @{\hskip 0\tabcolsep} |} \hline
    \multicolumn{1}{|c}{Hierarchy} & Bias $\hat{b}$ & Homophily $\hat{\lambda}$ & Years to parity (lowest level) & Years to parity (highest level) \\ \midrule[1pt]
    Nursing                                   & 0.228 & 7.74 & \cellcolor{red!25}$>200$ & \cellcolor{orange!25}134 \\ \hline
    Academic Chemistry               & 0.415 & 2.72 & \cellcolor{green!25}-17 & \cellcolor{lime!25}30 \\ \hline
    Academic Math/Stats              & 0.456 & 2.12 & \cellcolor{green!25}-24 & \cellcolor{red!25}$>200$ \\ \hline
    Academic Medicine (clinical)   & 0.474 & 1.75 & \cellcolor{green!25}-18 & \cellcolor{red!25}$>200$ \\ \hline
    Law                                          & 0.485 & 1.64 & \cellcolor{green!25}-24 & \cellcolor{red!25}$>200$ \\ \hline
    Academic Computer Science  & 0.489 & 4.49 & \cellcolor{red!25}$>200$ & \cellcolor{red!25}$>200$ \\ \hline
    Engineering Practice               & 0.496 & 4.82 & \cellcolor{red!25}$>200$ & \cellcolor{lime!25}9 \\ \hline
    Film*                                        & 0.497 & 5.76 & \cellcolor{green!25}-13 & \cellcolor{red!25}$>200$ \\ \hline
    Journalism                               & 0.499 & 3.47 & \cellcolor{lime!25}68 & \cellcolor{orange!25}108 \\ \hline
    Academic Medicine (science) & 0.503 & 2.37 & \cellcolor{green!25}-17 & \cellcolor{lime!25}51 \\ \hline
    Academic Physics                   & 0.506 & 4.51 & \cellcolor{red!25}$>200$ & \cellcolor{orange!25}101 \\ \hline
    Medical Practice                     & 0.523 & 0.34 & \cellcolor{green!25}-19 & \cellcolor{lime!25}26 \\ \hline
    Academic Biology                   & 0.544 & 2.66 & \cellcolor{green!25}-24 & \cellcolor{lime!25}7 \\ \hline
    Academic Psychology            & 0.562 & 3.58 & \cellcolor{red!25}$>200$ & \cellcolor{lime!25}13 \\ \hline
    Politics*                                  & 0.604 & 5.84 & \cellcolor{lime!25}68 & \cellcolor{lime!25}20 \\ \hline
    Academic Engineering           & 0.631 & 6.26 & \cellcolor{lime!25}20 & \cellcolor{lime!25}64 \\ \hline
\end{tabular}
     \caption{Best fit parameters for each hierarchy. Bias $b<0.5$ suggests that men may be disproportionately favored for promotion, while $b>0.5$ suggests that women may be disproportionately favored for promotion. Large homophily $\lambda$ suggests that gender is a salient factor when decided whether to apply for promotion. Years to parity indicates the predicted number of years beyond 2018 when gender fractionation will first be within 5\% of parity. Negative numbers indicate that gender parity within 5\% has already been reached, though may not remain within the gender parity band indefinitely. Times exceeding 200 years include predictions that gender parity will never be reached, though we hesitate to make predictions beyond that time frame.
*May not be a strict hierarchy: although producers hire directors, producers do not typically `promote' directors to producer positions. Likewise for politics.}
  \label{tab:results}
\end{center} 
\end{table}

Because our model does not require gender parity as an eventual outcome, our time-to-parity predictions are sensitive to small changes in parameter values, especially if the system is near a bifurcation. For instance, a small decrease in the bias parameter $b$ may shift time-to-parity from a small number of years to never reaching parity.

\begin{figure}[htb]
  \begin{center}
    \includegraphics[width=0.8\textwidth]{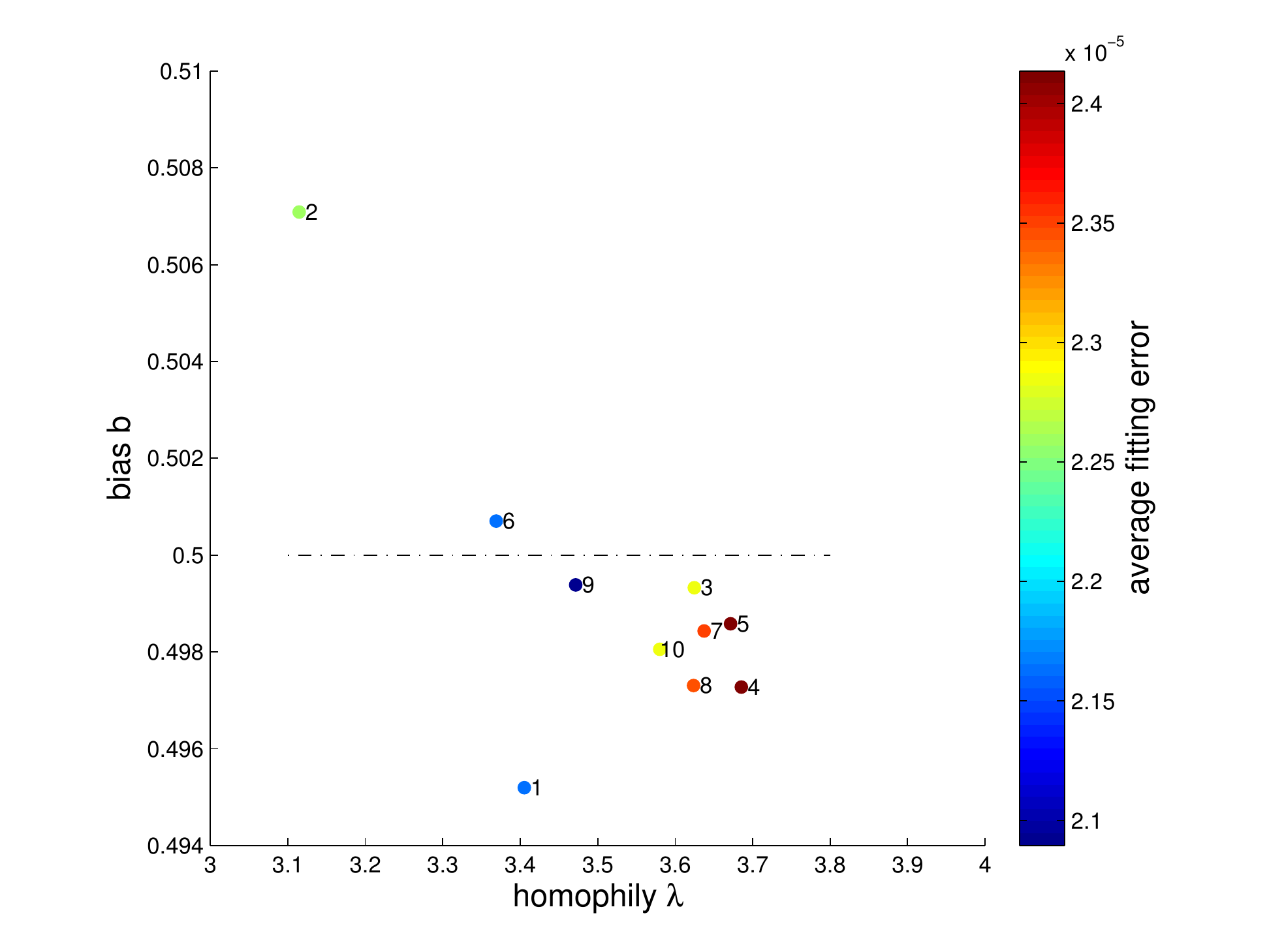} \\ \includegraphics[width=0.8\textwidth]{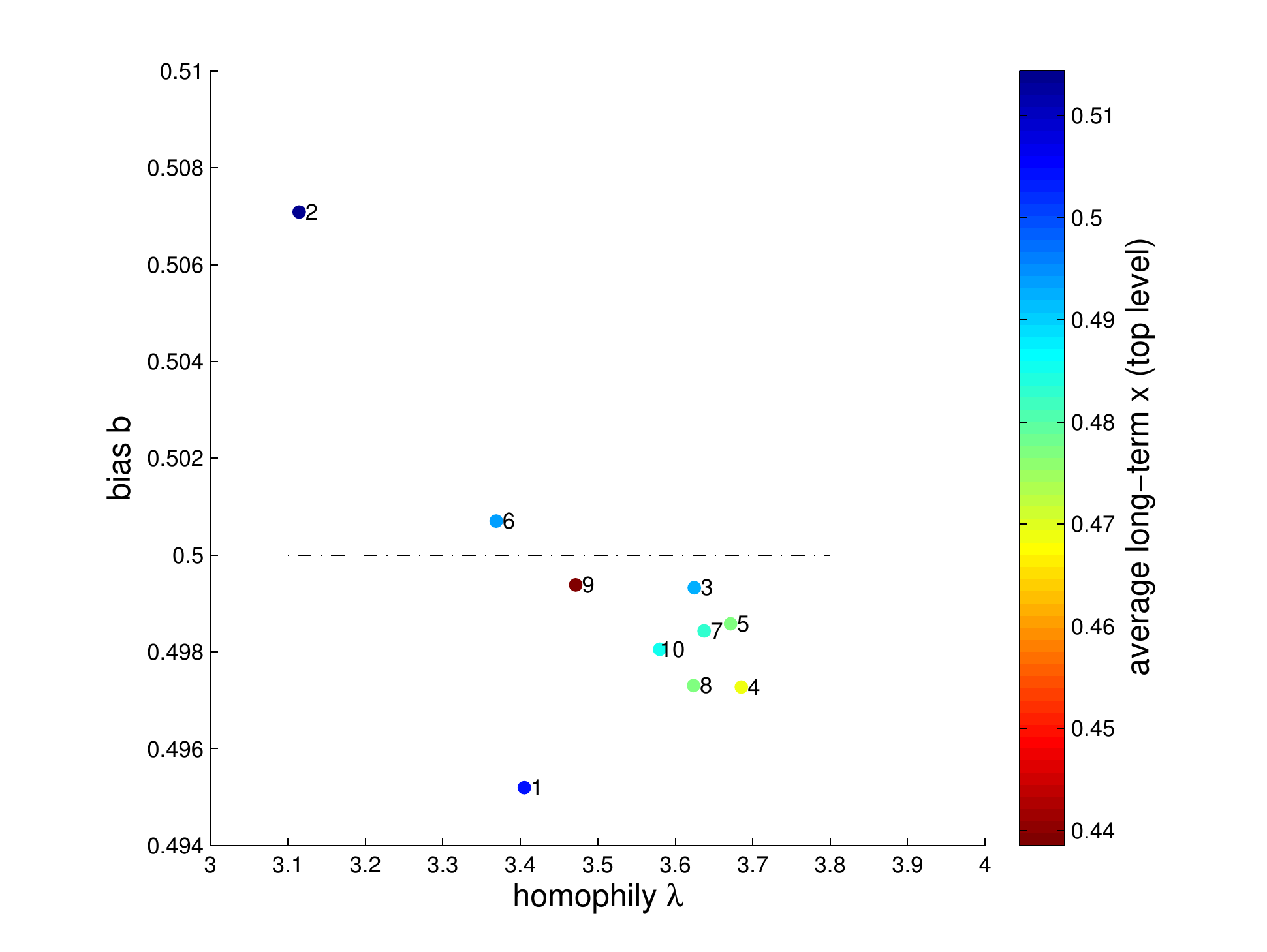}
  \end{center}
  \caption{Sensitivity of model predictions to changes in the model fitting algorithm for the journalism dataset. For ten random number generator seeds (right of points), the differences in error and predicted average long-term $x$ are small. For reference, the algorithm error tolerance is $1e-4$.} 
  \label{fig:senJour}
\end{figure}

\begin{figure}[htb]
  \begin{center}
    \includegraphics[width=0.8\textwidth]{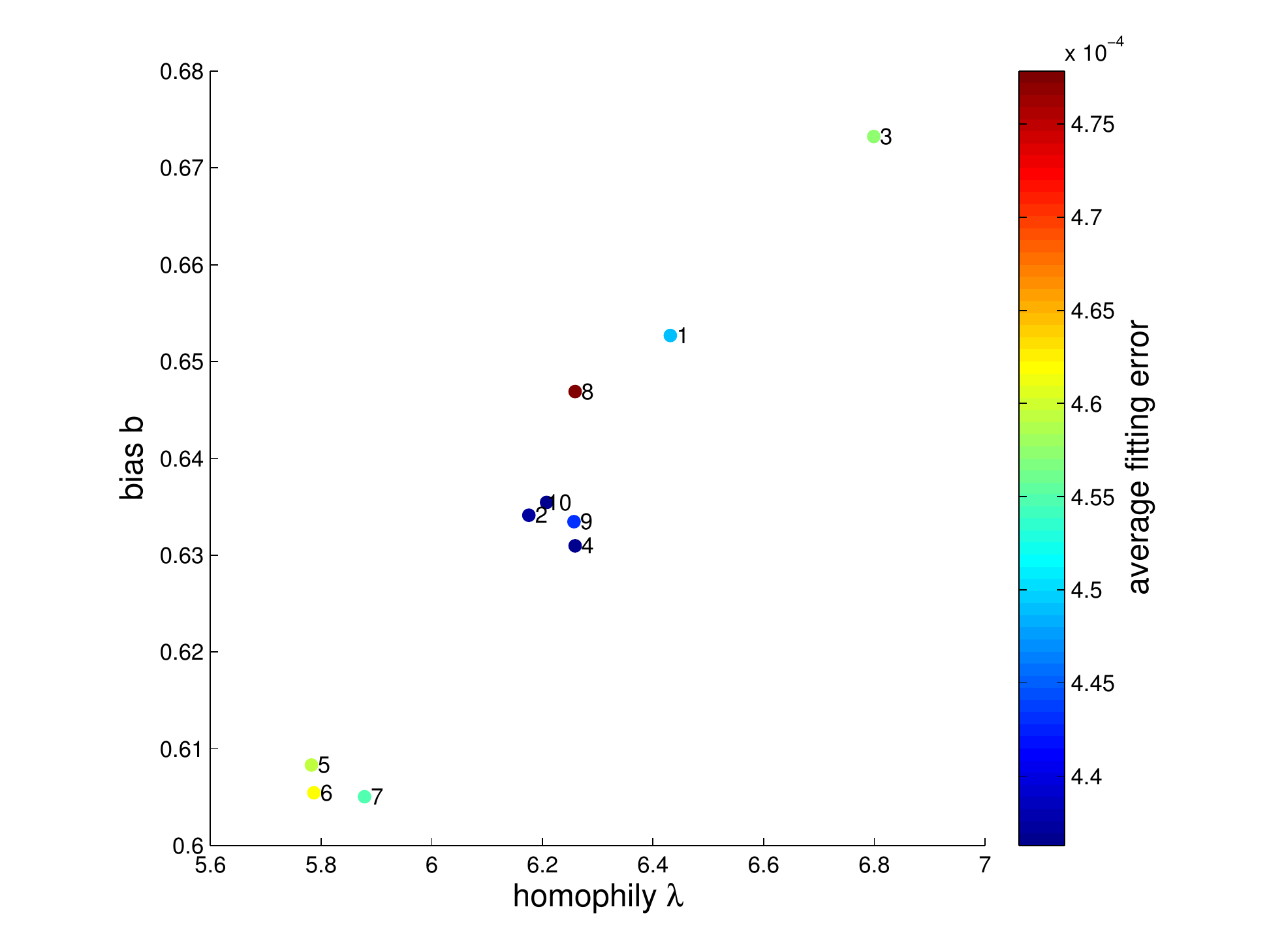} \\ \includegraphics[width=0.8\textwidth]{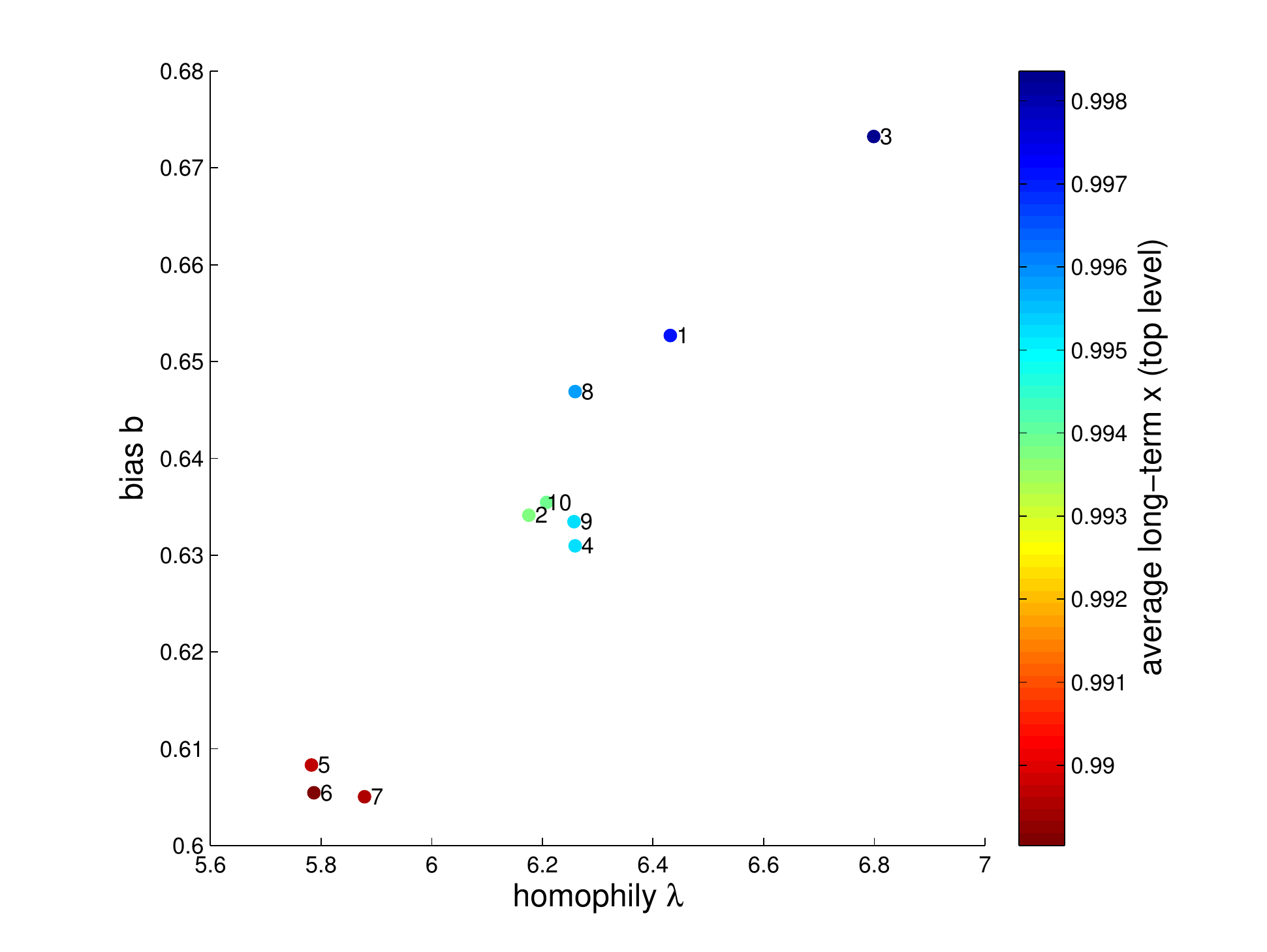}
  \end{center}
  \caption{Sensitivity of model predictions to changes in the model fitting algorithm for the academic engineering dataset. For ten random number generator seeds (right of points), the differences in error and predicted average long-term $x$ are small. For reference, the algorithm error tolerance is $1e-4$.}
  \label{fig:senEng}
\end{figure}

However, the average fractionation $x$ in each level is not sensitive to small changes in parameter values or to the model fitting procedure. See Figures \ref{fig:senJour} and \ref{fig:senEng} for two examples of fitting sensitivity.

\subsection{Comparison with Other Models}
To our knowledge, the model by Holman et al. \cite{holman2018gender} is the only other quantitative model capable of predicting time to gender parity. Their phenomenological model is
\begin{equation}
x_j = \frac{1}{2+c_j \exp{(-0.5 r_j t)}}  \,\,\,\,\, \text{for } 1\le j \le L,
\label{eq:holman}
\end{equation}
where $x_j$ is the fraction of people in the $j$th level who are women, $L$ is the number of levels, and $c_j$ and $r_j$ are fitting parameters. This model exhibits sigmoidal growth towards gender parity in most reasonable cases. Note that all hierarchy levels are independent of each other in this model.

We use our fitting algorithm to fit \eqref{eq:holman} to all datasets and compare the average sum of squared error between model and data. It is not possible to truly compare the models with our fitting algorithm because of differing initial parameter search ranges and time to algorithm convergence. The model by Holman et al. has an advantage that the search algorithm converges to a best fit quickly, and our model has an advantage of more fitting parameters.

With those caveats, we find that the average model errors differ by more than the fitting algorithm tolerance ($1e-4$) for seven out the 16 datasets. Among the seven datasets where our model errors differ, our model has a lower error in six of the seven cases. 

The model by Holman et al. better explains the dynamics seen in the dataset Academic Math/Statistics. Our model better explains the dynamics seen in the datasets Academic Engineering, Academic Computer Science, Academic Physics, Academic Biology, Academic Psychology, and Politics.

\end{document}